\documentclass[draftcls,onecolumn]{IEEEtran}
\usepackage{cite}
\ifCLASSINFOpdf
  \usepackage[pdftex]{graphicx}
\else
  \usepackage[dvips]{graphicx}
\fi
\usepackage[cmex10]{amsmath}
\usepackage{amssymb}
\usepackage{algorithm}
\usepackage{algorithmic}
\usepackage{array}
\usepackage[tight,footnotesize]{subfigure}
\usepackage{fixltx2e}
\usepackage{url}
\usepackage{nicefrac}
\usepackage{multirow}
\usepackage{color}
\usepackage{nicefrac}

\def\Skoric{\v{S}kori\'c }
\def\x{{\mathbf x}}
\def\y{{\mathbf y}}
\def\p{{\mathbf p}}

\def\cmax{c_{\mathsf{max}}}

\def\tmax{t_{\mathsf{max}}}
\def\pfp{P_{\mathsf{fp}}}
\def\pfn{P_{\mathsf{fn}}}
\def\pe{P_{\mathsf{e}}}
\def\sn{\sigma^2_{\mathsf{n}}}
\def\X{\mathcal{U}}
\def\code{\Xi}
\def\Acc{\mathcal{A}}

\def\xinn{\mathbf{x}_\mathsf{inn}}
\def\Tinn{\mathcal{T}_\mathsf{inn}}
\def\nt{n^{(t)}}
\def\pt{p^{(t)}}
\def\T{\mathcal{T}}
\def\Xt{\mathcal{U}^{(t)}}
\def\Xsi{\mathcal{U}_\mathsf{SI}}
\def\thetab{\boldsymbol{\theta}}
\def\rhob{\boldsymbol{\delta}}
\def\varphib{\boldsymbol{\varphi}}
\def\mub{\boldsymbol{\mu}}
\newcommand{\Prob}[1]{\mathbb{P}(#1)}
\newcommand{\Exp}[2]{\mathbb{E}_{#1}[#2]}
\newcommand{\nchoosek}[2]{{#1 \choose #2}}
\newtheorem{definition}{Definition}

\graphicspath{{figs/}}

\begin{document}
%
\title{Towards joint decoding of binary Tardos fingerprinting codes}
%
%
%

\author{Peter~Meerwald and Teddy~Furon
\thanks{P. Meerwald and T. Furon are with INRIA Rennes, France; e-mail: \{peter.meerwald, teddy.furon\}@inria.fr.}
\thanks{EDICS Category: WAT-FING}
}

%
%

\markboth{IEEE Transactions on Information Forensics and Security}%
{Meerwald \MakeLowercase{\textit{et al.}}: Towards joint Tardos decoding}
%


\maketitle

\begin{abstract}
The class of joint decoder of probabilistic fingerprinting codes is of utmost importance in theoretical papers to establish the concept of fingerprint capacity~\cite{Huang:2010fk,Amiri09a,Anthapadmanabhan08a}. However, no implementation supporting a large user base is known to date.
This article presents an iterative decoder which is, as far as we are aware of, the first practical attempt towards joint decoding. The discriminative feature of the scores benefits on one hand from the side-information of previously accused users, and on the other hand, from recently introduced universal linear decoders for compound channels~\cite{Abbe:2010yq}. Neither the code construction nor the decoder make precise assumptions about the collusion (size or strategy). The extension to incorporate soft outputs from the watermarking layer is straightforward. An extensive experimental work benchmarks the very good performance and offers a clear comparison with previous state-of-the-art decoders.
\end{abstract}


%
\IEEEpeerreviewmaketitle

\section{Introduction}
\label{sec:intro}
\IEEEPARstart{T}{raitor} tracing or active fingerprinting has witnessed 
a flurry of research efforts since the invention of the now well-celebrated Tardos codes~\cite{Tardos08a}. The codes of G.~Tardos are optimal in the sense that the code length $m$ necessary to fulfill the following requirements ($n$ users, $c$ colluders, probability of accusing at least one innocent below $\pfp$) has the minimum scaling in $\Omega(c^{2}\log n\pfp^{-1})$.

A first group of articles analyses such probabilistic fingerprinting codes from the viewpoint of information theory. They define the worst case attack a collusion of size $c$ can produce, and also the best counter-attack. The main achievement is a saddle point theorem in the game between the colluders and the code designer which establishes the concept of fingerprinting capacity $C(c)$~\cite{Huang:2010fk,Amiri09a,Anthapadmanabhan08a}. Roughly speaking, for a maximum size of collusion $c$, the maximum number of users exponentially grows with $m$ with an exponent equal to $C(c)$, to guarantee vanishing probabilities of error asymptotically as the code length increases. Sec.~\ref{Sec:Tardos} summarizes these elements of information theory. 

Our point of view is much more practical and signal processing oriented. Thanks to an appropriate watermarking technique, $m$ bits have been hidden in the distributed copies. At the time a pirated version is discovered, the content has been distributed to $n$ users. Our goal is to identify some colluders under the strict requirement that the probability of accusing innocents is below $\pfp$. It is clear that we are not in an asymptotic setup since $m$ and $n$ are fixed. The encoder and the decoder are not informed of the collusion size and its attack, therefore there is no clue whether the actual rate $R = m^{-1}\log_{2} n$ is indeed below capacity $C(c)$.

A second group of research works deals with decoding algorithms. Here, a first difficulty is to compute user scores that are as discriminative as possible. A second difficulty is to set a threshold such that one can reliably accuse users who are part of the collusion. These two steps are not easy since the decoder does not know the size and the attack of the collusion. Sec.~\ref{Sec:single} sums up the past approaches which are mainly based on single decoders. It also motivates our decoder based on compound channel theory and the use of a rare event estimator.

A third difficulty is to have a fast implementation of the accusation algorithm in order to face a large-scale set of users.
A main advantage of some fingerprinting schemes based on error-correcting codes is to offer an accusation procedure with runtime polynomial in $m$~\cite{Barg03a,Fernandez04a}.
In comparison, the well-known Tardos-\Skoric single decoder is an exhaustive search of complexity $O(nm)$~\cite{Tardos08a,Skoric2008:Symmetric}. Since in theory $n$ can asymptotically be in the order of $2^{mR}$, decoding of Tardos codes might be intractable. Again, we do not consider such a theoretical setup, but we pay attention to maintain an affordable decoding complexity for orders of magnitude met in practical applications.
 
Sec.~\ref{sec:joint} focuses on the iterative architecture of our joint decoder based on three primitives: channel inference, score computation, and thresholding. Its iterative nature stems from two key ideas: i) the codeword of a newly accused user is integrated as a side information for the next iterations, ii) joint decoding is manageable on a short list of suspects.    
Sec.~\ref{sec:soft} provides an extension to soft decoding.
In Sec.~\ref{sec:results} we present our experimental investigations with a comparison with related works for typical values of $(m,n)$. This shows the benefit of our decoder: better decoding performance with acceptable runtime in practical scenarios.

\section{Tardos code and the collusion model}\label{Sec:Tardos}
We briefly review the construction and some known facts about Tardos codes.
\subsection{Construction}
The binary code is composed of $n$ codewords of $m$ bits. The codeword
$\x_{j}=(x_{j}(1),\cdots,x_{j}(m))^T$ identifying
user $j \in \X =[n]$, where $[n]:=\{1,\dots,n\}$, is composed of $m$ binary symbols independently drawn at the code construction s.t.
$\Prob{x_{j}(i)=1}=p_i$, $\forall i\in[m]$. At initialization, the auxiliary variables $\{p_i\}_{i=1}^{m}$ are independent
and identically drawn according to distribution $f(p):[0,1]\rightarrow\mathbb{R}^{+}$. Both the code $\code=[\x_1,\dots,\x_n]$ and the auxiliary sequence $\p=(p_1,\ldots,p_m)^T$ must be kept as secret parameters.

\subsection{Collusion attack}\label{sub:coll}
The collusion attack or collusion channel describes the way the $c$ colluders $\mathcal{C}=\{j_{1},\ldots,j_{c}\}$ merge their binary codewords $\x_{j_{1}},\ldots,\x_{j_{c}}$ to forge the binary pirated sequence $\y$. It is usually modelled as a memoryless discrete multiple access channel, which is fair in the sense that all colluders participate equally in the forgery. This assumption comes from the fact that the worst case attacks are indeed memoryless for Tardos codes where symbols are generated independently, ~\cite[Lemma~3.3]{Moulin11a}. Moreover, in a detect-many scenario, there is no hope in identifying almost idle colluders if the attack is not fair~\cite[Lemma~3.2]{Moulin11a}.

This leads to a $2\times (c+1)$ probability transition matrix $[\Prob{Y|\varPhi}]$ where $\varPhi=\sum_{j\in\mathcal{C}}X_{j}$ is a random variable counting the number of `1' the colluders received out of $c$ symbols. A common parameter of the collusion attack on binary codes is denoted by the vector $\thetab_{c}=(\theta_{c}(0),\ldots,\theta_{c}(c))^{T}$ with $\theta_{c}(\varphi)=\Prob{Y=1|\varPhi=\varphi}$. The usual working assumption, so-called \emph{marking assumption}~\cite{Boneh98:fingerprinting}, imposes that $\theta_{c}(0)=1-\theta_{c}(c)=0$. The set of collusion attacks that $c$ colluders can lead under the marking assumption is denoted by $\Theta_{c}$:
\begin{equation}
\label{eq:DefTheta}
\Theta_{c} = \{\thetab\in[0,1]^{c+1}, \theta(0)=1-\theta(c)=0\}.
\end{equation}
Examples of attacks following this model are given, for instance, in~\cite{Furon2009:EM-decoding}.

\subsection{Accusation}
Denote $\Acc\subset\X$ the set of users accused by the decoder. The probability of false positive is defined by $\pfp =\Prob{\Acc\not\subset\mathcal{C}}$.
In practice, a major requirement is to control this feature so that it is lower than a given significance level.

In a detect-one scenario, $\Acc$ is either a singleton, or the empty set. A good decoder has a low probability of false negative defined by $\pfn=\Prob{\Acc=\emptyset}$. In a detect-many scenario, several users are accused, and a possible figure of merit is the number of caught colluders: $|\Acc\cap\mathcal{C}|$. In the literature, there exists a third scenario, so-called detect-all, where a false negative happens if at least one colluder is missed. This article only considers the first two scenarios.

\subsection{Guidelines from information theory}
\label{sub:ElemInfTheo}
This article does not pretend to any new theoretical contribution, but  presents some recent elements to stress guidelines when designing our practical decoder.
    
A \emph{single decoder} computes a score per user. It accuses users whose score is above a threshold (detect-many scenario) or the user with the biggest score above the threshold (detect-one scenario).
Under both scenarios and provided that the collusion is fair, the performance of such decoders is theoretically bounded by the achievable rate $R_{S}(f,\thetab_{c}) = I(X;Y|P,\thetab_{c}) = \Exp{P\sim f}{I(X;Y|p,\thetab_{c})}$~\cite[Th.~4.1]{Moulin11a}. A fundamental result is that, for a given collusion size $c$, there exists an equilibrium $(\breve{f}_{c,S},\breve{\thetab}_{c,S})$ to the max-min game between the colluders (who select $\thetab$) and the code designer (who selects $f$) as defined by $\max_{f}\min_{\thetab\in\Theta_{c}} R_{S}(f,\thetab)$ in~\cite[Th.~4]{Huang:2010fk}.

A \emph{joint decoder} computes a score per subset of $\ell \le c$ users and accuses the users belonging to subsets whose score is above a threshold or only the most likely guilty amongst these users. Under both scenarios and provided that the collusion is fair, the performance of such decoders is theoretically bounded by the achievable rate $R_{J}(f,\thetab_{c}) = \ell^{-1}I(\varPhi;Y|P,\thetab_{c}) = \ell^{-1}\Exp{P\sim f}{I(\varPhi;Y|p,\thetab_{c})}$~\cite[Th.~3.3]{Moulin11a}. $\varPhi$ denotes the random variable sum of the subset user symbols. Moreover, for a given collusion size $c$, there also exists an equilibrium $(\breve{f}_{c,J},\breve{\thetab}_{c,J})$ to the max-min game $\max_{f}\min_{\thetab\in\Theta_{c}} R_{J}(f,\thetab)$~\cite[Th.~4]{Huang:2010fk}.

Asymptotically, as $c\rightarrow +\infty$, both $\breve{f}_{c,J}$ and $\breve{f}_{c,S}$ converge to $f_{T}(p)=1/(\pi\sqrt{p(1-p)})$, the distribution originally proposed by G.~Tardos~\cite[Cor.~7]{Huang:2010fk}, and both $\min_{\thetab} R_{J}(f_{T},\thetab)$ and $\min_{\thetab} R_{S}(f_{T},\thetab)$ quickly approach the equilibrium value of the respective max-min game~\cite[Fig.~2]{Huang:2010fk}. Yet, 
the code designer needs to bet on a collusion size $c^{\prime}$ in order to use the optimal distribution $\breve{f}_{c^{\prime},S}$ (or $\breve{f}_{c^{\prime},J}$ if the decoder is joint). 
Integer $c^{\prime}$ plays the role of a desired security level.

Despite the division by $\ell$ in the expression of $R_{J}(f,\thetab)$, it appears that $R_{S}(f,\thetab)\leq R_{J}(f,\thetab),\,\forall\thetab$~\cite[Eq.~(3.4)]{Moulin11a}. This tells us that a joint decoder is theoretically more powerful than a single decoder. However, a joint decoder needs to compute $O(n^{\ell})$ scores since there are $\nchoosek{n}{\ell}$ subsets of size $\ell$. This complexity is absolutely intractable for large-scale applications even for a small $\ell$. This explains why, so far, joint decoders were only considered theoretically to derive fingerprinting capacity. Our idea is that there is no need to consider all these subsets since a vast majority is only composed of innocent users. Our decoder iteratively prunes out users deemed as innocents and considers the subsets over the small set of remaining suspects. 

This iterative strategy results in a decoder which is a mix of single and joint decoding. Unfortunately, it prevents us from taking advantage of the game theory theorems mentioned above. We cannot find the optimal distribution $f$ and the worst collusion attack against our decoder. Nevertheless, our decoder works with any distribution $f$ under some conditions stated in Sec.~\ref{Sec:single}.
For all these reasons, the experiments of Sec.~\ref{sec:results} are done with the most common Tardos distribution $f_{T}$.


M.~Fernandez and M.~Soriano proposed an iterative accusation process of an error correcting code based fingerprinting scheme~\cite{Fernandez04a}.  Each iteration takes advantage of the codewords of colluders already identified in the previous iterations. The same idea is possible with Tardos probabilistic fingerprinting code. This is justified by the fact that the side information $\Delta$, defined as the random variable sum of the already identified colluder symbols, increases the mutual information:  $I(\varPhi;Y|P,\thetab_{c})\leq I(\varPhi;Y|P,\thetab_{c},\Delta)$.
Indeed, side information helps more than joint decoding as proved by~\cite[Eq.~(3.3)]{Moulin11a}.

The above guidelines can be summarized as follows: use the continuous Tardos distribution $f_{T}$ for code construction, integrate the codewords of accused users as side information and finally use a joint decoder on a short list of suspects.

\section{A single decoder based on compound channel theory and rare event analysis}\label{Sec:single}
This section first reviews some single decoders and presents new decoders based on compound channel theory and rare event analysis. The first difficulty is to compute a score per user such that the colluders are statistically well separated from the innocents scores. The second difficulty is to set a practical threshold such that the probability of false positive is under control.

Detection theory tells us that the score given by the Log-Likelihood Ratio (LLR):
\begin{equation}
s_{j} = \sum_{i=1}^{m}\log\frac{\Prob{y(i)|x_{j}(i),\thetab_{c}}}{\Prob{y(i)|\thetab_{c}}},
\label{eq:MAP}
\end{equation}
is optimally discriminative in the Neyman-Pearson sense to decide the guiltiness of user $j$. Yet, the LLR needs the knowledge of the true collusion attack $\thetab_{c}$ which prevents the use of this optimal single decoder in practical settings. Some papers proposed a so-called `Learn and Match' strategy using the LLR score tuned on an estimation $\hat{\thetab}$ of the attack channel~\cite{Furon2009:EM-decoding}. Unfortunately, a lack of identifiability obstructs a direct estimation from $(\y,\p)$ (see Sec.~\ref{sub:ApplSingle}).
Indeed, the estimation is sound only if $c$ is known, and if the number of different values taken by $p$ is bigger\footnote{This is the case in this article since we opt for the continuous Tardos distribution $f_{T}$.} or equal than $c-1$: $\Prob{Y=1|\thetab,p}$ is a polynomial in $p$ of degree at most $c$ (see~\eqref{eq:GenericProba} with $u=0$ and $v=0$) going from point $(0,0)$ to $(1,1)$, we need $c-1$ more points to uniquely identify this polynomial.
To overcome this lack of information about $c$, an Expectation-Maximization (E.-M.) approach has been proposed but it is not satisfactory since it does not scale well with the number of users~\cite{Furon2009:EM-decoding}. Moreover, the setting of the threshold was not addressed.

On the other hand, there are decoders that do not adapt their score computation to the collusion.
This is the case of the score computation originally proposed by G. Tardos~\cite{Tardos08a}, and later-on improved by B. \Skoric {\it et al.}~\cite{Skoric2008:Symmetric}. It has an invariance property: its statistics, up to the second order, do not depend on the collusion attack channel $\thetab$, but only on the collusion size $c$~\cite{Furon2008:On-the-design}. Thanks to this invariance, whatever the collusion attack is, there exists a threshold $\tau$ guaranteeing a probability of false positive below $\pfp$ while keeping the false negative away from 1 provided that the code is long enough, {\it i.e.} $m = \Omega(c^{2}\log n\pfp^{-1})$.
However, there is a price to pay: the scores are clearly less discriminative than the LLR.  

Some theoretical papers~\cite[Sec. V]{Somekh-Baruch:2005uq}~\cite[Sec.~5.2]{Moulin11a} promote another criterion, so-called `universality', for the design of decoders. The performance (usually evaluated as the achievable rate or the error exponent) when facing a collusion channel $\thetab_{c}$ should not be lower than the performance against the worst attack $\thetab^{\star}_{c}$. In a sense, it is a clear warning to the `Learn and Match' strategy. Suppose that $\thetab_{c}\neq\thetab^{\star}_{c}$ and that, for some reasons, the estimation of the collusion attack is of poor quality. In any case, a mismatch between $\hat{\thetab}$ and $\thetab_{c}$ should not ruin the performance of the decoder to the point it is even lower than what is achievable under the worst attack $\thetab^{\star}_{c}$. The above cited references~\cite{Somekh-Baruch:2005uq,Moulin11a} recommend the single universal decoder based on the empirical mutual information $I(\x;\y|\p)$ (or empirical equivocation for joint decoder). The setting of the threshold depends on the desired error exponent of the false positive rate. Therefore, it is valid only asymptotically.  

To summarize, there have been two approaches: adaptation or non-adaptation to the collusion process. The first class is not very well grounded since the estimation of the collusion is an issue and the impact of a mismatch has to be studied. The second approach is more reliable, but with a loss of discrimination power compared to the optimal LLR. The next sections presents two new decoders belonging to both approaches based on the compound channel theory.

\subsection{Some elements on compound channels}
Recently, in the setup of digital communication through compound channels, E.~Abbe  and L.~Zheng~\cite{Abbe:2010yq} proposed universal decoders which are linear, {\it i.e.} in essence very simple. This section summarizes this theory and the next one proposes two applications for Tardos single decoders.

A compound channel is a set $\mathcal{S}$ of channels, say discrete memoryless channels $X\in\mathcal{X}\rightarrow Y\in\mathcal{Y}$ defined by their probability transition matrix $W_{\theta}=[\Prob{Y|X,\theta}]$ parameterized by $\theta\in\Theta$. The coder shares a code book $\code=\{\x_{j}\}_{j=1}^{n}\in\mathcal{X}^{m\times n}$ with the decoder. Its construction is assumed to be a random code realization from a provably good mass distribution $P_{X}$. After receiving a channel output $\y\in\mathcal{Y}^{m}$, a decoder computes a score per codeword $\x_{j}$, $j\in[n]$, and yields the message associated with the codeword with the biggest score. The decoder is linear if the score has the following structure:
\begin{equation}
s_{j} = \sum_{i=1}^{m} d(x_{j}(i),y(i)),
\end{equation}
with $d(\cdot,\cdot):\mathcal{X}\times\mathcal{Y}\rightarrow\mathbb{R}$. For instance, score~\eqref{eq:MAP}, so-called MAP decoder in digital communications~\cite{Abbe:2010yq}, is linear with $d(x,y)=\log(\Prob{y|x,\theta}/\Prob{y|\theta})$. However, in the compound channel setup, the decoder does not know through which channel
of $\mathcal{S}$ the codeword has been transmitted, and therefore it cannot rely on the MAP.

We are especially interested in two results. First, if $\mathcal{S}$ is {\it one-sided} (see Def.~\ref{def:OneSided} below), then the MAP decoder tuned on the worst channel $W_{\theta^{\star}}$ is a linear universal decoder~\cite[Lemma~5]{Abbe:2010yq}. If $\mathcal{S}=\bigcup_{k=1}^{K}\mathcal{S}_{k}$ with $K$ finite and $\mathcal{S}_{k}$ one-sided $\forall k\in[K]$, then the following {\it generalized} linear decoder is universal~\cite[Th.~1]{Abbe:2010yq} and the score of a codeword is the maximum of the $K$ MAP scores tuned on the worst channel $W_{\theta_{k}^{\star}}$ of each $\mathcal{S}_{k}$:
\begin{equation}
s_{j} = \max_{k\in[K]} \sum_{i=1}^{m} \log\frac{\Prob{y(i)|x_{j}(i),\theta_{k}^{\star}}}{\Prob{y(i)|\theta_{k}^{\star}}}.
\end{equation}

\begin{definition}[One-sided set, Def.~3~of~\cite{Abbe:2010yq}]\label{def:OneSided}
A set $\mathcal{S}$ is one-sided with respect to an input distribution $P_{X}$
\begin{itemize}
\item if the following minimizer is unique:
\begin{equation}
W_{\theta^{\star}} = \arg \min_{\theta\in\mathsf{cl}(\Theta)} \mathcal{I}(P_{X},\theta),
\end{equation}
with $\mathcal{I}(P_{X},\theta)$ the mutual information $I(X;Y)$ with $(X,Y)\sim P_{X}\circ W_{\theta}$ (where $P\circ W$ denotes the joint distribution with $P$ the distribution of $X$ and $W$ the conditional distribution), and $\mathsf{cl}(\Theta)$ the closure of $\Theta$,
\item and if, $\forall \theta\in\Theta$,
\begin{equation}\label{eq:DefOneSided}
\begin{aligned}
D(P_{X}\circ W_{\theta}||P_{X}\times P_{Y,\theta^{\star}})\geq & D(P_{X}\circ W_{\theta}||P_{X}\circ W_{\theta^{\star}}) + \\
& D(P_{X}\circ W_{\theta^{\star}}||P_{X}\times P_{Y,\theta^{\star}}).
\end{aligned}
\end{equation}
with $D(\cdot||\cdot)$ the Kullback-Leibler distance, $P_{Y,\theta}$ the marginal of $Y$ induced by $P_{X}\circ W_{\theta}$, and $P_{X}\times P_{Y,\theta}$ the product of the marginals.
\end{itemize}
\end{definition}

\subsection{Application to single Tardos decoders}\label{sub:ApplSingle}
Contrary to the code construction phase, it is less critical at the decoding side to presume that the real collusion size $c$ is less or equal to a given parameter $\cmax$. This parameter can be set to the largest number of colluders the fingerprinting code can handle with a reasonable error probability knowing  $(m,n)$. Another argument is that this assumption is not definitive. If the decoding fails because the assumption does not hold true, nothing prevents us to re-launch decoding with a bigger $\cmax$. 
Let us assume $c\leq\cmax$ in the sequel.

A first application of the work~\cite{Abbe:2010yq} is straightforward:  The collusion channel belongs to the set $\bigcup_{k=2}^{\cmax}\Theta_{k}$ as defined~\eqref{eq:DefTheta}, and thanks to~\cite[Lemma 4]{Abbe:2010yq} each convex set $\Theta_{k}$ is one-sided. According to~\cite[Th.~1]{Abbe:2010yq}, the decoder based on the following score is universal:
\begin{equation}
s_{j} = \max_{k\in[2,\ldots,\cmax]} \sum_{i=1}^{m}\log\frac{\Prob{y(i)|x_{j}(i),\thetab_{k,f_{T}}^{\star}}}{\Prob{y(i)|\thetab_{k,f_{T}}^{\star}}},
\label{eq:compound}
\end{equation}
where $\thetab_{k,f_{T}}^{\star} = \arg\min_{\Theta_{k}} R_{S}(f_{T},\thetab)$, $\forall k\in[2,\ldots,\cmax]$. This decoder does not adapt its score computation to the collusion attack.

The second application is more involved as the lack of identifiability turns to our advantage. The true collusion channel $\thetab_{c}$ has generated data $\y$ distributed as $\Prob{y|p,\thetab_{c}}$.  Let us define the class $\mathcal{E}(\thetab_{c})=\{\tilde{\thetab}| \Prob{y|p,\tilde{\thetab}}=\Prob{y|p,\thetab_{c}},\,\forall (y,p)\in\{0,1\}\times[0,1]\}$. Thanks to~\cite[Prop. 3]{Furon2009:Worst-WIFS}, we know that $\mathcal{E}(\thetab_{c})$ is not restricted to the singleton $\{\thetab_{c}\}$ since for any $c^{\prime}>c$ there exists one $\tilde{\thetab}_{c^{\prime}}\in\mathcal{E}(\thetab_{c})$. This holds especially for $\cmax$.
Asymptotically with the code length, the consistent Maximum Likelihood Estimator (MLE) parameterized on $\cmax$, as defined in~\eqref{eq:MLE}, yields an estimation $\hat{\thetab}_{\cmax}\approx\tilde{\thetab}_{\cmax}\in\mathcal{E}(\thetab_{c})$ with increasing accuracy. This estimation is not reliable because $c\neq\cmax$ a priori. Therefore, we prefer to refer to $\hat{\thetab}_{\cmax}$ as a collusion inference rather than a collusion estimation, and the scoring uses this inference as follows:
\begin{equation}
s_{j} = \sum_{i=1}^{m}\log\frac{\Prob{y(i)|x_{j}(i),\hat{\thetab}_{\cmax}}}{\Prob{y(i)|\hat{\thetab}_{\cmax}}}.
\label{eq:single}
\end{equation}

Suppose that the MLE tuned on $\cmax$ provides a perfect inference $\hat{\thetab}_{\cmax}=\tilde{\thetab}_{\cmax}$, we then succeed to restrict the compound channel to the discrete set $\mathcal{E}_{\cmax}(\thetab_{c})$ which we define as the restriction of $\mathcal{E}(\thetab_{c})$ to collusions of size $\tilde{c}\leq\cmax$. Appendix~\ref{appendix} shows that $\mathcal{E}_{\cmax}(\thetab_{c})$ is one-sided, and its worst attack is indeed $\tilde{\thetab}_{\cmax}$. Lemma 5 of~\cite{Abbe:2010yq} justifies the use of the MAP decoder~\eqref{eq:MAP} tuned on $\hat{\thetab}_{\cmax}$. Its application leads to a more efficient decoder since $R_{S}(f_{T},\tilde{\thetab}_{\cmax})\geq R_{S}(f_{T},\thetab_{\cmax,f_{T}}^{\star})$. This decoder pertains to the approach based on score adaptation, with the noticeable advantages: it is better theoretically grounded and it is far less complex than the iterative E.-M. decoder of~\cite{Furon2009:EM-decoding}.  

\begin{figure}[t]
\begin{center}
\subfigure[\textit{Worst-Case} Attack]{\includegraphics[width=0.6\columnwidth]{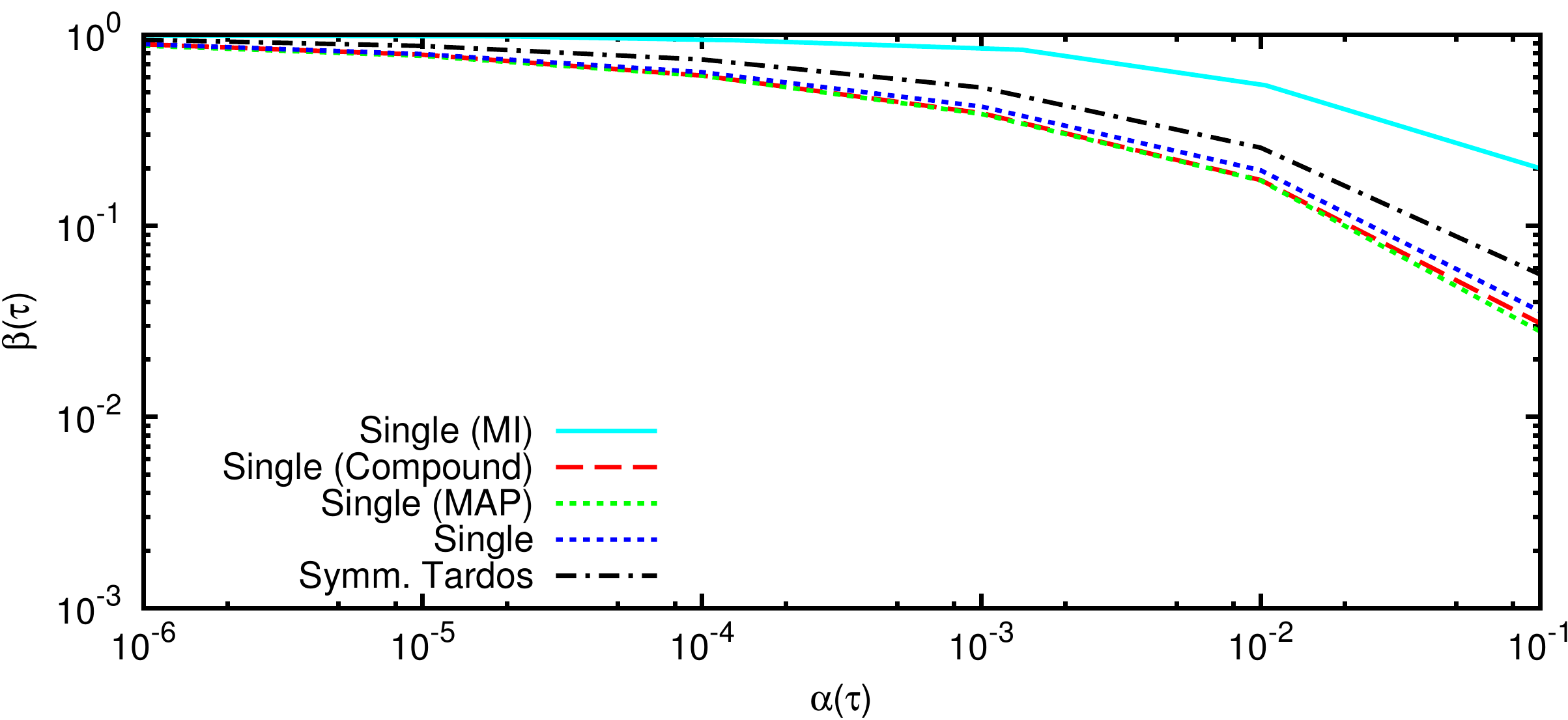}}
\subfigure[\textit{Majority} Attack]{\includegraphics[width=0.6\columnwidth]{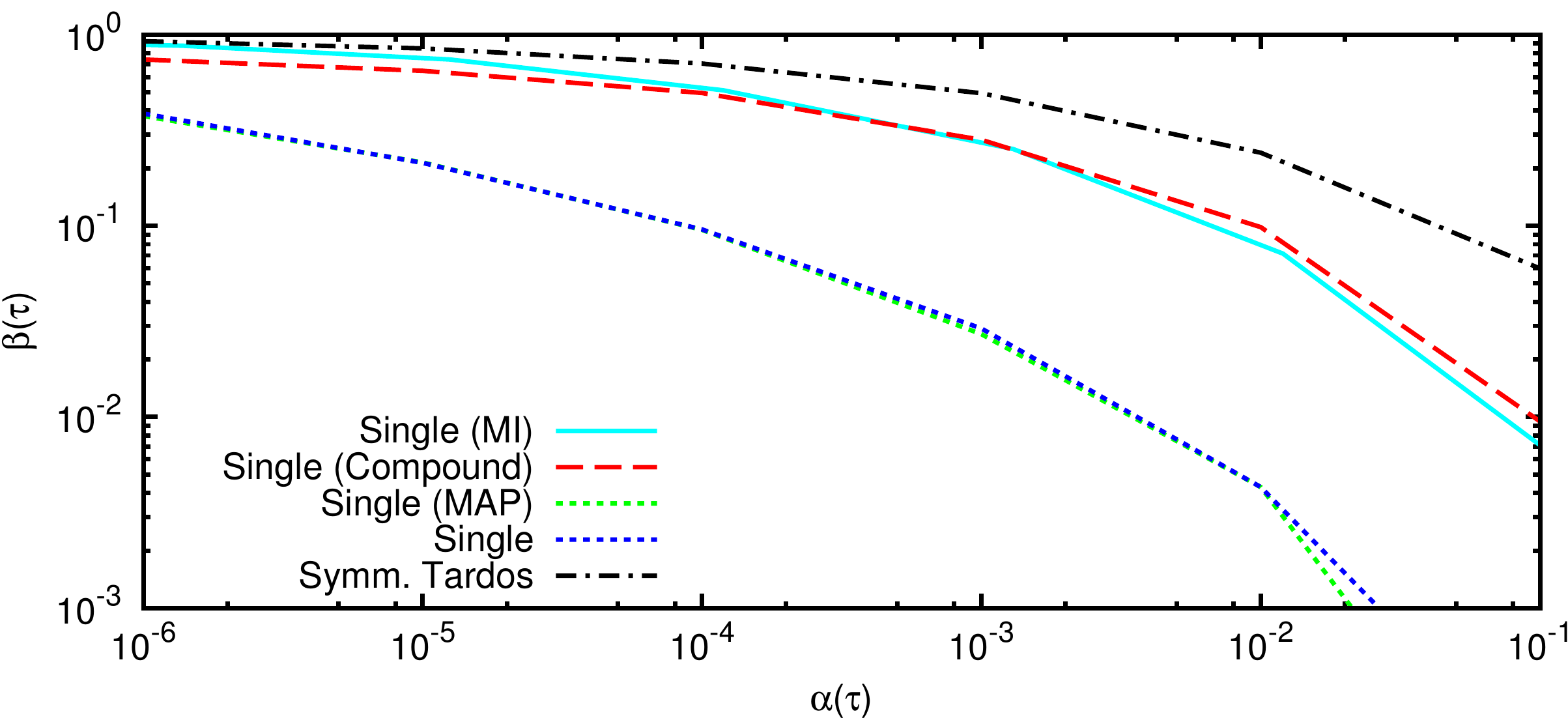}}
\end{center}
\caption{ROC plot for several decoders; $m=512$, $c=5$, $\cmax=8$. Single (MI)
is the decoder based on empirical mutual information~\cite{Moulin11a},
Single (Compound) relates to~\eqref{eq:compound}, Single (MAP) is~\eqref{eq:MAP}, Single is the LLR on $\hat{\thetab}_{\cmax}$~\eqref{eq:single}, and Symm. Tardos is the symmetric version of the G.~Tardos scores proposed by B.~\Skoric {\it et al.} in~\cite{Skoric2008:Symmetric}.\label{fig:roc}}
\end{figure}
Figure~\ref{fig:roc} illustrates the Receiver Operating Characteristics (ROC) per user for the single decoders discussed so far with $m=512$ and $c=5$ colluders performing \textit{worst-case} ({\it i.e.} minimizing $R_{S}(f_{T},\thetab)$ over $\Theta_{5}$) and \textit{majority} attack ($\theta_{5,\mathsf{maj}}=(0,0,0,1,1,1)^{T}$). For this figure, the false positive $\alpha(\tau)$ and the false negative $\beta(\tau)$ are defined \emph{per user} as follows:
\begin{eqnarray}
\alpha(\tau)&=&\Prob{s(\xinn,\y,\p)>\tau},\\
\beta(\tau)&=&\Prob{s(\mathbf{x}_{j_{1}},\y,\p)\leq\tau},
\end{eqnarray}
where $\xinn$ is a random variable denoting the codeword of an innocent user and $\mathbf{x}_{j_{1}}$, the codeword of the first colluder.
The \textit{single} decoder is tuned on the collusion inference $\hat{\thetab}_{\cmax}$ (with $\cmax=8$) and performs almost as good as the MAP decoder having knowledge of $\thetab$. The ROC of the symmetric Tardos score is invariant w.r.t. the collusion attack. The generalized linear decoder of~\eqref{eq:compound} denoted \textit{compound} takes little advantage of the fact that the majority attack is much milder than the worst attack.
For a fair comparison, the single decoder based on the empirical mutual information \cite{Moulin11a} assumes a Tardos distribution uniformly quantized to $10$ bins; better results (yet still below the \textit{single} decoder) can be obtained when tuned to the optimal discrete distribution for $c=5$ colluders \cite{Huang09c}.

The similarities between compound channel and fingerprinting has been our main inspiration, however some differences prevent any claim of optimality.
First, in the compound channel problem, there is a unique codeword that has been transmitted, whereas in fingerprinting, $\y$ is forged from $c$ codewords like in a multiple access channel. Therefore, the derived single decoders are provably good for chasing a given colluder (detect-one scenario), but they might not be the best when looking for more colluders (detect-many scenario).
The second difference is that the decoder should give up when not confident enough rather than taking the risk of being wrong in accusing an innocent. The setting of a threshold is clearly missing for the moment.

\subsection{Rare event analysis}
\label{sub:RarEveA}
This section explains how we set a threshold $\tau$ in accordance with the required $\pfp$ thanks to a rare event analysis. Our approach is very different than~\cite{Somekh-Baruch:2005uq}\cite{Moulin11a}\cite{Amiri09a}\cite{Tardos08a} where a theoretical development either finds a general threshold suitable when facing a collusion of size $c$, or equivalently, where it claims a reliable decision when the rate is below the capacity which depends on $c$. Our threshold does not need the value of $c$ but it only holds for a given couple $(\p,\y)$ and a known $n$. Once these are fixed, the scoring $s_{j}=s(\x_{j},\y,\p)$ is a deterministic function from $\{0,1\}^{m}$ to $\mathbb{R}$. Since the codewords of the innocent users are i.i.d. and $c\ll n$, we have:
\begin{equation}
\begin{aligned}
\pfp&=1 - (1-\Prob{s(\xinn,\y,\p)>\tau})^{n-c} \\
&\approx n \cdot \Prob{s(\xinn,\y,\p)>\tau}.
\end{aligned}
\end{equation}
The number of possible codewords can be evaluated as the number of typical sequences, {\it i.e.} in the order of $2^{m\Exp{P\sim f}{h_{b}(p)}}$, with $h_{b}(p)$ the entropy in bits of a Bernoulli random variable $B(p)$. $\Exp{P\sim f_{T}}{h_{b}(p)}\approx 0.557$ bits, which leads to a far bigger number of typical sequences than $n$ (say $m\geq 300$ and $n\leq 10^{8}$ in practice). This shows that plenty of codewords have not been created when a pirate copy is found. Therefore, we consider them as occurrences of $\xinn$ since we are sure that they have not participated in the forgery of $\y$. The idea is then to estimate $\tau$ s.t. $\Prob{s(\xinn,\y,\p)>\tau}=n^{-1}\pfp$ thanks to a Monte Carlo simulation with newly created codewords.

The difficulty lies in the order of magnitude. Some typical requirements are $n\approx 10^{6}$ and $\pfp=10^{-4}$, hence the estimation of $\tau$ corresponding to a probability as small as $10^{-10}$. This is not tractable with a basic Monte Carlo on a regular computer. However, the new estimator based on rare event analysis proposed in~\cite{Guyader:2011kx} performs remarkably fast within this range of magnitude. It produces $\hat{\tau}$ and a $C$-\% confidence interval\footnote{We are $C$-\% sure that the true $\tau$ lies in this interval.} $[\tau^{-},\tau^{+}]$. In our decoder, we compare the scores to $\tau^{+}$ ({\it i.e.} a pessimistic estimate of $\tau$) to ensure a total false positive probability lower than $\pfp$. Last but not least, this approach works for any single decoder.

\section{Iterative, Joint decoding algorithm}
\label{sec:joint}

\begin{figure}[t]
\begin{center}
\includegraphics[width=0.8\columnwidth]{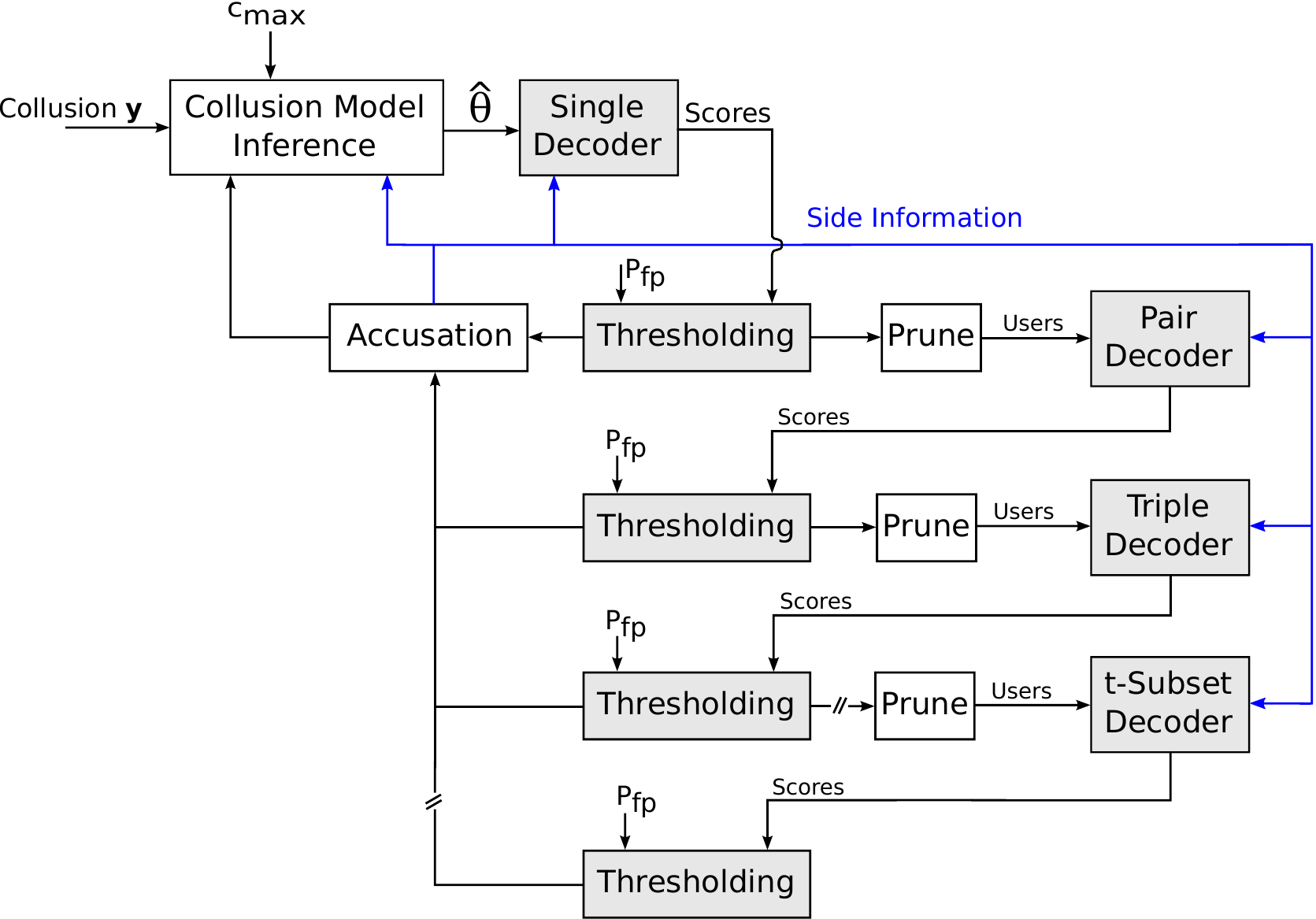}
\end{center}
\caption{Overview of the iterative, side-informed joint Tardos fingerprint decoder.\label{fig:overview}}
\end{figure}

This section extends the single decoder based on the collusion inference $\thetab_{\cmax}$ towards joint decoding, thanks to the guidelines of Sec.~\ref{sub:ElemInfTheo}. Preliminary results about these key ideas were first presented in~\cite{Meerwald11b} and~\cite{Meerwald11c}. A schematic overview of the iterative, joint decoder is shown in Fig.~\ref{fig:overview}. 

\subsection{Architecture}
The first principle is to iterate the score computation and include users accused in previous iterations as side-information to build a more discriminative test. Let $\Xsi = \emptyset$ denote the initially empty set of accused users. In each iteration we aim at identifying a (possibly empty) set of users $\mathcal{A} =  \{j \in \X \setminus \Xsi | s_j > \tau \}$ and then update $\Xsi$ with $\mathcal{A}$. 

Second, we additionally compute scores for subsets of $t$ users of $\X \setminus \Xsi$, $t \le \cmax$. Obviously, there are $\nchoosek{|\X \setminus \Xsi|}{t}$ such subsets. As $n$ is large, enumerating and computing a score for each subset is intractable even for small $t$. The idea here is to find a restricted set $\Xt \subseteq \X \setminus \Xsi$ of $\nt=|\Xt|$ users that are the most likely to be guilty and keep $\pt=\nchoosek{\nt}{t}$ approximately constant and within our computation resources. We gradually reduce $\nt$ by pruning out users who are unlikely to be colluder when going from single ($t=1$) decoding, to pair ($t=2$) decoding, etc. If $\nt=O(n^{\nicefrac{1}{t}})$, then score computation of $t$-subsets over the restricted user set is within $O(n)$ just like for the single decoder.

Initially, the joint pair-decoder starts with the list of users ranked by the scores derived from the single decoder in decreasing order, {\it i.e.} the top-ranked user is most likely to be a colluder. Later on, the joint $t$-subset decoder produces a new list of scores computed from subsets of $t$ users which -- according to theoretical results \cite{Amiri09a,Moulin11a} -- are more discriminative as $t$ increases. 
Denote $\T^\diamond \subseteq \Xt$ the $t$-subset of users with the highest score. Our algorithm tries to accuse the most likely colluder within $\T^\diamond$, and, if successful, updates $\Xsi$ and continues with the single decoder.
If no accusation can be made, the algorithm generates a new list of suspects $\X^{(t+1)}$ based on the ranking of joint scores that is fed to the subsequent $t+1$ joint decoding stage. 

In the detect-one scenario, iteration stops after the first accusation. We restrict the subset size to $t \le \tmax$, with $\tmax=5$. This is not a severe limitation as for moderately large $c$, the decoding performance advantage of the joint decoder quickly vanishes \cite{Moulin11a}.
In the detect-many scenario, iteration stops when $|\Xsi| \ge \cmax$ or $t$ reaches $\min(\tmax,\cmax-|\Xsi|)$ and no further accusation can be made. The set $\Xsi$ then contains the user indices to be accused. Alg.~\ref{algo:iterative-joint} illustrates the architecture of the accusation process for the catch-many scenario.

The next sections describe the score computation, the accusation of a user and the inference of the collusion process in more details.

\begin{algorithm}
\caption{Iterative Joint Tardos Decoder.\label{algo:iterative-joint}}
\begin{algorithmic}[1]
\REQUIRE $\mathbf{y}$, $\code$, $\mathbf{p}$, $\cmax$, $\tmax \le \cmax$, $\nt$, $\pfp$

\STATE $\X \gets \{j | 1 \le j \le n\}$, $\Xsi \gets \emptyset$
\REPEAT
  \STATE $t \gets 1$
  \STATE $\hat{\thetab}_{\cmax} \gets \texttt{infere}(\y, \mathbf{p}, \Xsi, \cmax)$
  \STATE $\textbf{W} \gets \texttt{weights}(\mathbf{y}, \mathbf{p}, \hat{\thetab}_{\cmax}, \Xsi$) 
  \STATE $\textbf{s} \gets \texttt{scores}(\X \setminus \Xsi, \code, \textbf{W})$
  \STATE $\tau \gets \texttt{threshold}(\mathbf{p}, \mathbf{W}, n^{-1}\pfp)$
  \STATE $\mathcal{A} \gets \{j \in \X \setminus \Xsi | s_j > \tau\}$
  \WHILE{$\mathcal{A}=\emptyset$ \AND $t<\tmax$}
    \STATE $t \gets t+1$
    \STATE $\Xt \gets \{j \in \X \setminus \Xsi | s_j > \texttt{top}(\textbf{s}, \nt) \}$
    \STATE $\textbf{W} \gets \texttt{weights}(\mathbf{y}, \mathbf{p}, \hat{\thetab}_{\cmax}, \Xsi$)
    \STATE $\textbf{s} \gets \texttt{scores}(\nchoosek{\Xt}{t}, \code, \textbf{W})$
    \STATE $\tau \gets \texttt{threshold}(\mathbf{p}, \mathbf{W}, \nchoosek{n}{t}^{-1}\pfp, t)$
    \STATE $\T^\diamond \gets \underset{\T \in \Xt}{\arg\max} ~ s_\T$
    \IF{$s_{\T^\diamond} > \tau$}
      \FORALL{$j \in \T^\diamond$ \AND \textbf{while} $\mathcal{A}=\emptyset$}
        \STATE $\mathbf{W} \gets \texttt{weights}(\mathbf{y}, \mathbf{p}, \hat{\thetab}_{\cmax}, \Xsi \cup \{\T^\diamond \setminus j\})$
        \STATE $\tau' \gets \texttt{threshold}(\mathbf{p},\mathbf{W},n^{-1}\pfp)$
        \STATE $\mathcal{A} \gets \{j | \texttt{score}(j,\code,\mathbf{W}) > \tau'\}$
      \ENDFOR
    \ENDIF
  \ENDWHILE
  \STATE $\Xsi \gets \Xsi \cup \mathcal{A}$
\UNTIL $\mathcal{A}=\emptyset$ \OR $|\Xsi| \ge \cmax$
\RETURN $\Xsi$
\end{algorithmic}
\end{algorithm}

\subsection{Score computation}

For a $t$-subset $\T$, the accusation is formulated as a hypothesis test based on the observations $(\textbf{y},\{\mathbf{x}_j\}_{j\in\T})$ to decide between $\mathcal{H}_0$ (all $j \in \T$ are innocent) and $\mathcal{H}_1$ (all $j \in \T$ are guilty). The score is just the LLR tuned on the inference $\hat{\thetab}_{\cmax}$ of the collusion process.

All these sequences are composed of independent random variables thanks to the code construction and the memoryless nature of the collusion. Moreover, the collusion only depends on the number of symbol `1' present in the codewords of a subset.
Therefore, denote by $\rhob$ and $\varphib$ the accumulated codewords corresponding to $\Xsi$ and $\mathcal{T}$: $\rhob = \sum_{j\in \Xsi}\mathbf{x}_j$ and $\varphib=\sum_{j \in \T}\mathbf{x}_j$. We have $\forall i\in[m],\, 0\leq\delta(i)\leq n_{\mathsf{SI}}$ and $0\leq\varphi(i)\leq t$. 
Thanks to the linear structure of the decoder, the score for a subset $\T$ of $t$ users is simply
\begin{equation}
s_\T = \sum^m_{i=1} W(\varphi(i),i),
\label{eq:scoring}
\end{equation}
where the $(t+1)\times m$ weight matrix $\mathbf{W}$ is pre-computed from $(\y,\p)$ taking into account the side information $\Xsi$ so that $\forall (\varphi,i) \in \{0,\dots,t\}\times\{1,\dots,m\}$:
\begin{equation}
W(\varphi,i)=\log \frac{\Prob{y(i) | (\varphi,t),(\delta(i),n_{\mathsf{SI}}),p(i),\hat{\thetab}_{\cmax}}}{\Prob{y(i) |(\delta(i),n_{\mathsf{SI}}),p(i),\hat{\thetab}_{\cmax}}}.
\label{eq:sT}
\end{equation}
For indices s.t. $y(i)=1$, both the numerator and the denominator share a generic formula, $P(\varphi(i)+\delta(i),t+n_{\mathsf{SI}},p(i),\hat{\thetab}_{\cmax})$ and $P(\delta(i),n_{\mathsf{SI}},p(i),\hat{\thetab}_{\cmax})$ respectively, with 
\begin{equation}
\begin{aligned}
P(u,v,p,\hat{\thetab}_{\cmax})= & \sum_{k=u}^{\cmax-v+u}\hat{\theta}_{\cmax}(k) \cdot \\
& \nchoosek{\cmax-v}{k-u}p^{k-u}(1-p)^{\cmax-v-k+u}. 
\end{aligned}
\label{eq:GenericProba}
\end{equation}
In words, this expression gives the probability that $y=1$ knowing that the symbol `1' has been distributed to users with probability $p$, the collusion model $\hat{\thetab}_{\cmax}$, and the identity of $v$ colluders who have $u$ symbols `1' and $v-u$ symbols `0'.
For indices s.t. $y(i)=0$ in~\eqref{eq:sT}, the numerator and the denominator need to be `mirrored': ($P\rightarrow 1-P$). 

At iterations based on the single decoder: $t=1$ and $\varphib=\mathbf{x}_{j}$ for user $j$. If nobody has been deemed guilty so far, then $\delta(i)=n_{\mathsf{SI}}=0,\,\forall i\in[m]$. This score is defined if $t+n_{\mathsf{SI}}\leq \cmax$. Therefore, for a given size of side-information, we cannot conceive a score for subsets of size bigger than $\cmax-n_{\mathsf{SI}}$. This implies that in the detect-many scenario, the maximal number of iterations depends on how fast $\Xsi$ grows.

\subsection{Ranking users within a subset and joint accusation}

Let $\T^\diamond$ denote the $t$-subset with the highest score. We accuse one user in $\T^\diamond$ only if $s_{\T^\diamond}>\tau$. Let $\Tinn$ denote a subset composed of innocent users. Using rare event analysis, $\tau$ is estimated s.t. $\Prob{s(\{\x_{j}\}_{j \in \Tinn},\y,\p)>\tau}=\nchoosek{n}{t}^{-1}\pfp$.
This thresholding operation ensures that $\T^\diamond$ contains at least one colluder with a very high probability.

In order to rank and accuse the most probable traitor in $\T^\diamond$, we record for each user $j \in \Xt$ the subset leading to that user's highest score:
\begin{equation}
\T^\diamond_j = \underset{\T}{\arg\max} \{s_\T | j \in \T\}.
\end{equation}

We can count how often each user $j$ appears in the recorded subsets $\{\T^\diamond_j\}_{j \in \Xt}$ and denote this value $a_j$. Finally, for a given $\T$, the users $j_k \in \T$ can be arranged s.t. $a_{j_1} \ge a_{j_2} \ge \dots \ge a_{j_t}$ to establish a ranking of users per subset.\footnote{This detail is omitted in Alg.~\ref{algo:iterative-joint} but necessary for procedure \texttt{top()}.}

To accuse a user $j \in \T^\diamond$, we check if the single score $s(\x_j,\y,\p,\Xsi \cup \{\T^\diamond \setminus j\}) > \tau'$ with $\tau'$ s.t. $\Prob{s(\xinn,\y,\p,\Xsi \cup \{\T^\diamond \setminus j\})>\tau'}=n^{-1}\pfp$. This method is suggested in~\cite[Sec. 5.3]{Moulin11a}.

\subsection{Inference of the collusion process}
\label{sec:inference}
The MLE is used to infer about the collusion process:
\begin{equation}
\hat{\thetab}_{\cmax} = \arg\max_{\thetab \in \Theta_{\cmax}} \log \Prob{\y|\p,\Xsi,\thetab}.
\label{eq:MLE}
\end{equation}
Whenever a user is deemed guilty, it is added to side-information and we re-run the parameter estimation to refine the collusion inference.

\section{Soft Decoding under AWGN attack}
\label{sec:soft}

\begin{figure}[t]
\begin{center}
\includegraphics[width=0.8\columnwidth]{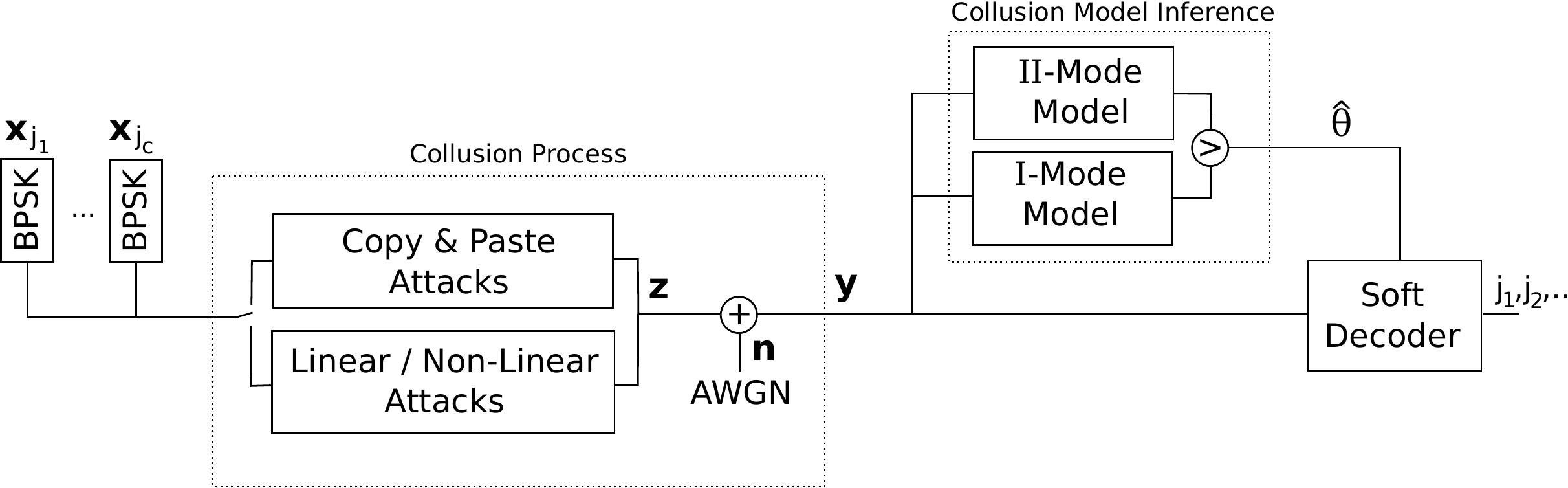}
\end{center}
\caption{Attack channel and collusion model inference.\label{fig:attack}}
\end{figure}

\begin{figure}[t]
\begin{center}
\includegraphics[width=0.8\columnwidth]{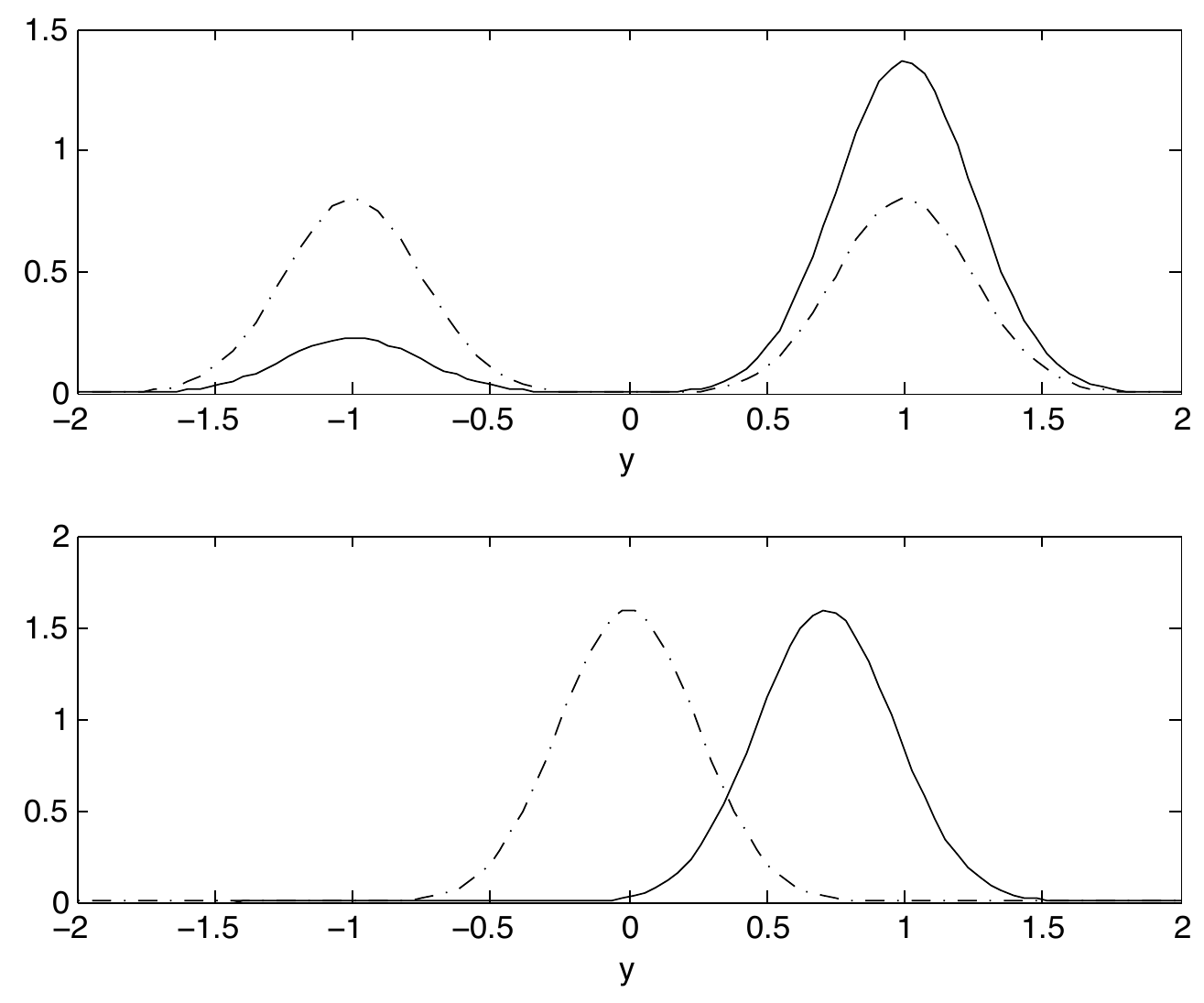}
\end{center}
\caption{Examples of pdf $\theta_{7}(y'|6)$ for the two models with $\sn=0.25$: [top] two modes ($II$) with (solid) the interleaving attack ($\theta(\varphi)=\varphi/c$) and (dashed) the coin-flip attack ($\theta(\varphi)=1/2$ for $0<\varphi<c$) ; (bottom) one mode ($I$) with (solid) averaging attack ($\mu(\varphi)=2c^{-1}\varphi-1$) and (dashed) set to 0 attack ($\mu(\varphi)=0$ for $0<\varphi<c$). \label{fig:mode}}
\end{figure}

The marking assumption is an unrealistic restriction for traitor tracing with multimedia content as the colluders are not limited to the copy-and-paste strategy for each symbol. They can merge the samples of their content versions (audio samples, pixels, DCT coefficients, etc.) in addition to traditional attempts to compromise the watermark.
This may result in erroneously decoded symbols or erasures from the watermarking layer. Relaxing the marking assumption leads to several approaches such as the combined digit model~\cite{Skoric09a}~\cite[Sec.~4]{Perez-Freire2009:Blind} and soft-decision decoding schemes~\cite{Kuribayashi10e, Schaathun08b}. This section extends the capability of our joint decoder to this latter case, replacing the probability transition $2\times (c+1)$ matrix $[\Prob{Y|\varPhi}]$ (see Sec.~\ref{sub:coll}) by $c+1$ probability density functions $\{\theta_{c}(y|\varphi)\}_{\varphi=0}^{c}$.

It is challenging if not impossible to exhibit a model encompassing all the merging attacks while being relevant for a majority of watermarking techniques. Our approach as sketched in Fig.~\ref{fig:attack} is pragmatic. The sequence $\y'\in\mathbb{R}^{m}$ is extracted from the pirated copy, with modulation $y'(i)=2y(i)-1$ if the signal is perfectly watermarked with binary symbol $y(i)$. To reflect the merging attack, the colluders forge values $z(i)\in[-1,1]$ and add noise: $y'(i)=z(i)+n(i)$ with $n(i)\sim\mathcal{N}(0,\sn)$.
This would be the case, for instance, for a spread spectrum watermarking where a symbol is embedded per block of content with an antipodal modulation of a secret carrier~\cite{Kuribayashi10e,Jourdas09a}.

The colluders have two strategies to agree on $\mathbf{z}$. In a first strategy, they collude according to the marking assumption ({\it i.e.} they copy-and-paste one of their samples) and add noise: $\mathbf{z}\in\{-1,1\}^{m}$ and the probability that $z=1$ is given by the components of $\thetab_{c}$.   
\begin{equation}
\theta^{(II)}_{c}(y'|\varphi)=\left(\theta_{c}(\varphi) e^{-\frac{(y'-1)^2}{2\sn}} + (1-\theta_{c}(\varphi)) e^{\frac{(y'+1)^2}{2\sn}}\right) / \sqrt{2\pi\sn}
\label{eq:mode2}
\end{equation}
Except for $\varphi\in\{0,c\}$, the pdfs have a priori two modes (hence the superscript $II$). This model is parameterized by $(\thetab,\sn)$. 

In a second strategy, the colluders select $z(i)=\mu(\varphi(i))\in[-1,1]$:
\begin{equation}
\theta^{(I)}_{c}(y'|\varphi)=e^{-\frac{(y'-\mu(\varphi))^2}{2\sn}} / \sqrt{2\pi\sn}.
\label{eq:mode1}
\end{equation}
An equivalent of the marking assumption would impose that $\mu(0)=-1$ and $\mu(c)=1$. The pdfs have a unique mode (hence the superscript $I$). This model is parameterized by $(\mub,\sn)$. Fig.~\ref{fig:mode} gives some examples of such pdfs.

A simple approach, termed \textit{hard} decision decoding in the sequel, consists in first thresholding $\y'$ (to quantize $y'(i)$ into $0$ if $y'(i) < 0$ and $1$ otherwise), and then employ the collusion process inference of Sec.~\ref{sec:inference} on the hard outputs.
Our \textit{soft} decision decoding method resorts to the noise-aware models \eqref{eq:mode2} and \eqref{eq:mode1} and sets
\begin{equation} 
\hat{\thetab}_{\cmax}=\underset{\thetab \in \{\hat{\thetab}^{(II)}_{\cmax}, \hat{\thetab}^{(I)}_{\cmax} \}}{\arg\max} \Prob{\y|\p,\Xsi,\thetab}.
\end{equation}
Notice that models $I$ and $II$ share the same number of parameters, therefore, there is no risk of over-fitting.  


\section{Experimental Results}
\label{sec:results}

We implemented the Tardos decoders in C++\footnote{Source code is available at \url{http://www.irisa.fr/texmex/people/furon/src.html}.}. Single and joint score computation is implemented efficiently using pre-computed lookup tables, cf. \eqref{eq:scoring} and \eqref{eq:sT}, and aggregation techniques described in \cite{Meerwald11b}. For a code length of $m=1024$ more than $10^6$ single and about $10^5$ joint scores, respectively, can be computed per second on single core of a regular Intel Core2 $2.6$~GHz CPU.
To control the runtime, the joint decoders are confined to $5$-subset decoding ($\tmax=5$) and $\pt \approx 4.5\cdot10^{6}$ computed subsets per joint decoding stage.
An iterative decoding experiment can be executed on a PC within a couple of minutes, given enough memory, see \cite{Meerwald11c} for details.
To experimentally verify the false-positive rate controlled by rare-event analysis, up to $3\cdot10^{4}$ tests per parameter setting have been performed on a cluster of PCs.

First, we first compare the performance of the proposed decoders under marking assumption. Finally, we lift this unrealistic restriction and turn to a more practical assessment using soft-decision decoding.

Unless explicitly noted, the terms \textit{single} and \textit{joint} decoder refer to the decoders conditioned on the inference of the collusion process $\hat{\thetab}_{\cmax}$, cf. \eqref{eq:single} and \eqref{eq:scoring}.
Further, we consider the MAP decoders assuming knowledge of $\thetab_c$ and the compound channel decoder, cf. \eqref{eq:compound}, tuned on the worst-case attack $\thetab^\star_{k,f_T}$, $\forall k \in [2,\dots,\cmax]$. As a baseline for a performance comparison, we always include symmetric Tardos score computation \cite{Skoric2008:Symmetric} with a threshold controlled by rare-event analysis (see Sec.~\ref{sub:RarEveA}).

\subsection{Decoding performance under marking assumption}

\subsubsection{Detect-one scenario}

Here the aim is to catch at most one colluder -- this is the tracing scenario most commonly considered in the literature. We compare our \textit{single} and \textit{joint} decoder performance against the results provided by Nuida~{\it et~al.}~\cite{Nuida09a} (which are the best as far as we know) and, as a second reference, the symmetric Tardos decoder. 

\begin{figure}[t]
\begin{center}
\includegraphics[width=0.8\columnwidth]{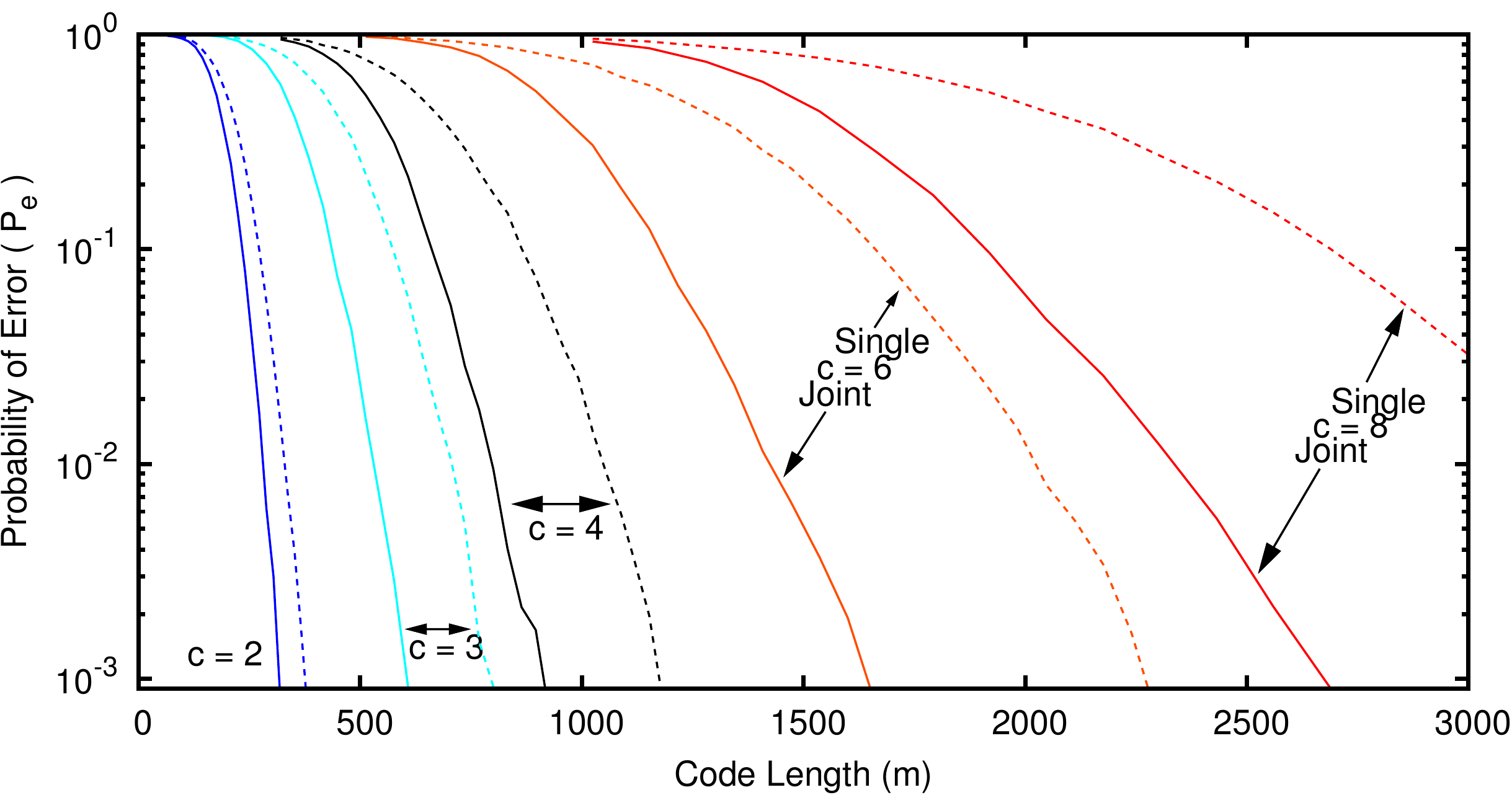}
\end{center}
\caption{Code length vs. $\pe$ for $n=10^6$ users and different number of colluders performing \textit{worst-case} attack against a single decoder; $\cmax=8$.\label{fig:nuida-catch-one}}
\end{figure}

The experimental setup considers $n=10^{6}$ users and $c \in \{2,3,4,6,8\}$ colluders performing \textit{worst-case} attack \cite{Furon2009:Worst-WIFS} against a single decoder. In Fig.~\ref{fig:nuida-catch-one}, we plot the empirical probability of error $\pe = \pfp + \pfn$ obtained by running $10^4$ experiments for each setting versus the code length $m$. 
The false-positive error is controlled by thresholding based on rare-event simulation, $\pfp=10^{-3}$, which is confirmed experimentally. Evidently, for a given probability of error, the \textit{joint} decoder succeeds in reducing the required code length over the \textit{single} decoder, especially for larger collusions.

Table~\ref{tbl:nuida-catch-one} compares the code length to obtain an error rate of $\pe=10^{-3}$ for our proposed Tardos decoders and the symmetric Tardos decoder with the results reported by Nuida~{\it et~al}.~\cite{Nuida09a} under marking assumption.
Except for $c=2$, the proposed decoders can substantially reduce the required code length and the \textit{joint} decoder improves the results of the \textit{single} decoder.
Note that Nuida's results give analytic code length assuming a particular number of colluders for constructing the code while our results are experimental estimates based on worst-case attack against a single decoder and without knowing $c$ (subject to $c \le \cmax=8$). Results with $c$ known are provided in \cite{Meerwald11c} and show a slightly better performance: the required code length of the \textit{joint} decoder is then slightly shorter than Nuida's code in case $c=2$.

\begin{table}[t]
\begin{center}
\caption{Code length comparison for the detect-one scenario: $n=10^{6}$, Worst-Case attack against a single decoder, $\pe=10^{-3}$. \label{tbl:nuida-catch-one}}
\begin{tabular}{c|cc|cc}
\multirow{2}{1cm}{\centering Colluders ($c$)} & \multirow{2}{1.5cm}{\centering Nuida \textit{et~al.} \cite{Nuida09a}} & 
\multirow{2}{1.5cm}{\centering Symm. Tardos} &
\multicolumn{2}{c}{Proposed ($\cmax=8$)}  \\
& & & Single & Joint \\
\hline 
2 & $253$ & $\sim 416$ & $\sim 368$ & $\sim 304$  \\
3 & $877$ & $\sim 864$ & $\sim 776$ & $\sim 584$  \\
4 & $1454$ & $\sim 1472$ & $\sim 1152$ & $\sim 904$  \\
6 & $3640$ & $\sim 2944$ & $\sim 2304$ & $\sim 1616$  \\
8 & $6815$ & $\sim 5248$ & $\sim 3712$ & $\sim 2688$  
\end{tabular}
\end{center}
\end{table}

\subsubsection{Detect-many scenario}

We now consider the more realistic case where the code length $m$ is fixed and the false-negative error rate is only a minor concern\footnote{A tracing schemes rightly accusing a colluder half of the time might be enough to dissuade dishonest users.} while the false-positive probability is critical to avoid an accusation of an innocent.  The aim is to identify as many colluders as possible. 

\begin{figure}[t]
\begin{center}
\subfigure[\textit{Worst-Case} Attack against Single Decoder\label{fig:many-wca}]{\includegraphics[width=0.6\columnwidth]{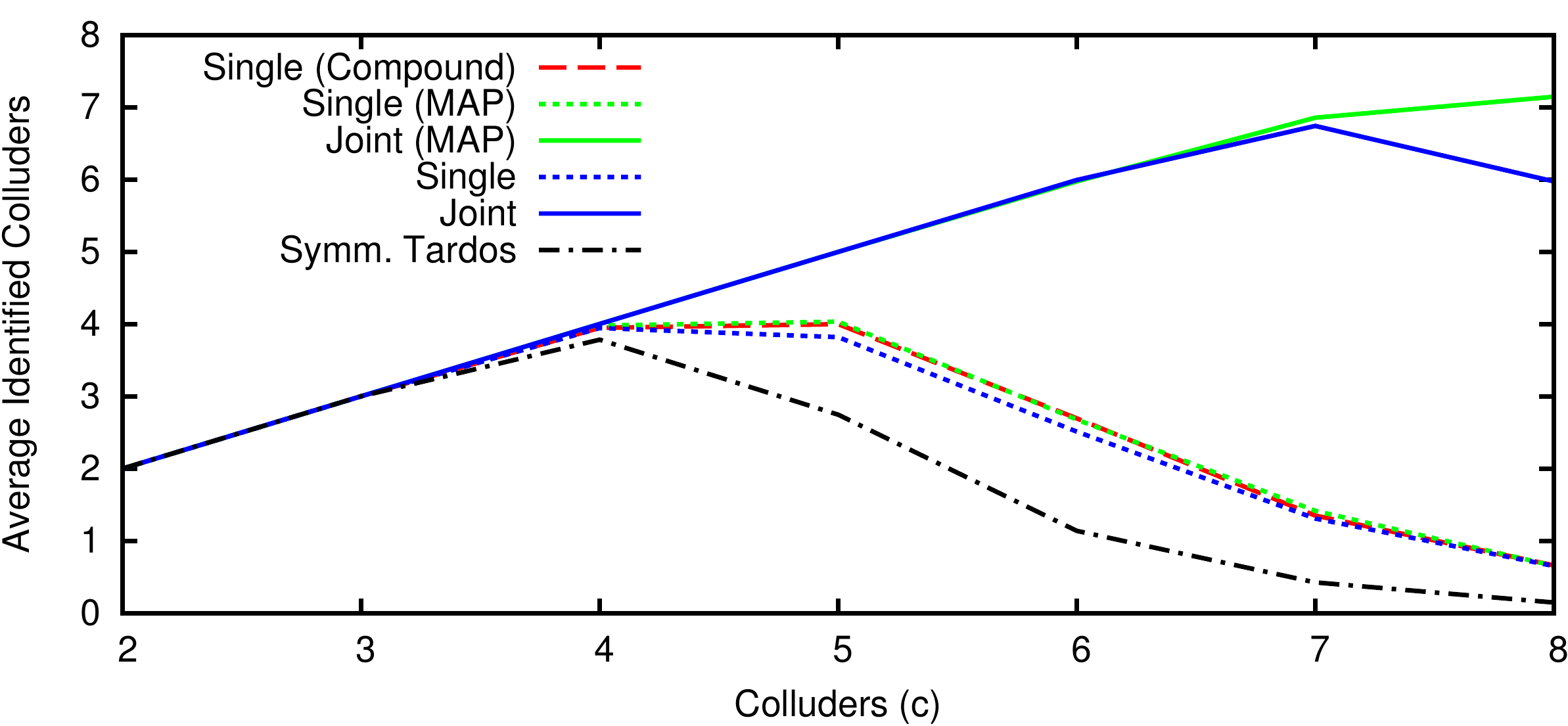}}
\subfigure[\textit{Worst-Case} Attack against Joint Decoder\label{fig:many-wcaj}]{\includegraphics[width=0.6\columnwidth]{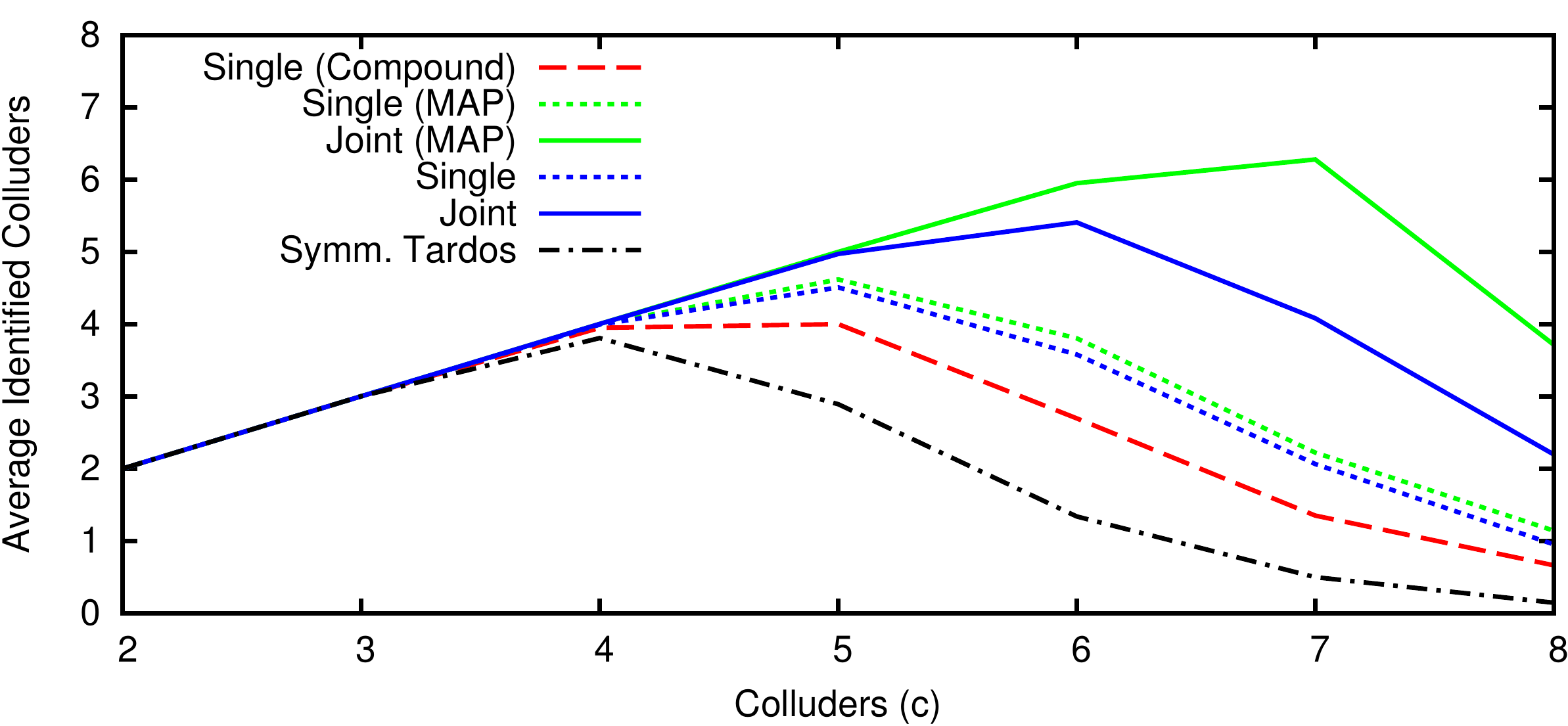}}
\subfigure[\textit{Interleaving} Attack\label{fig:many-unif}]{\includegraphics[width=0.6\columnwidth]{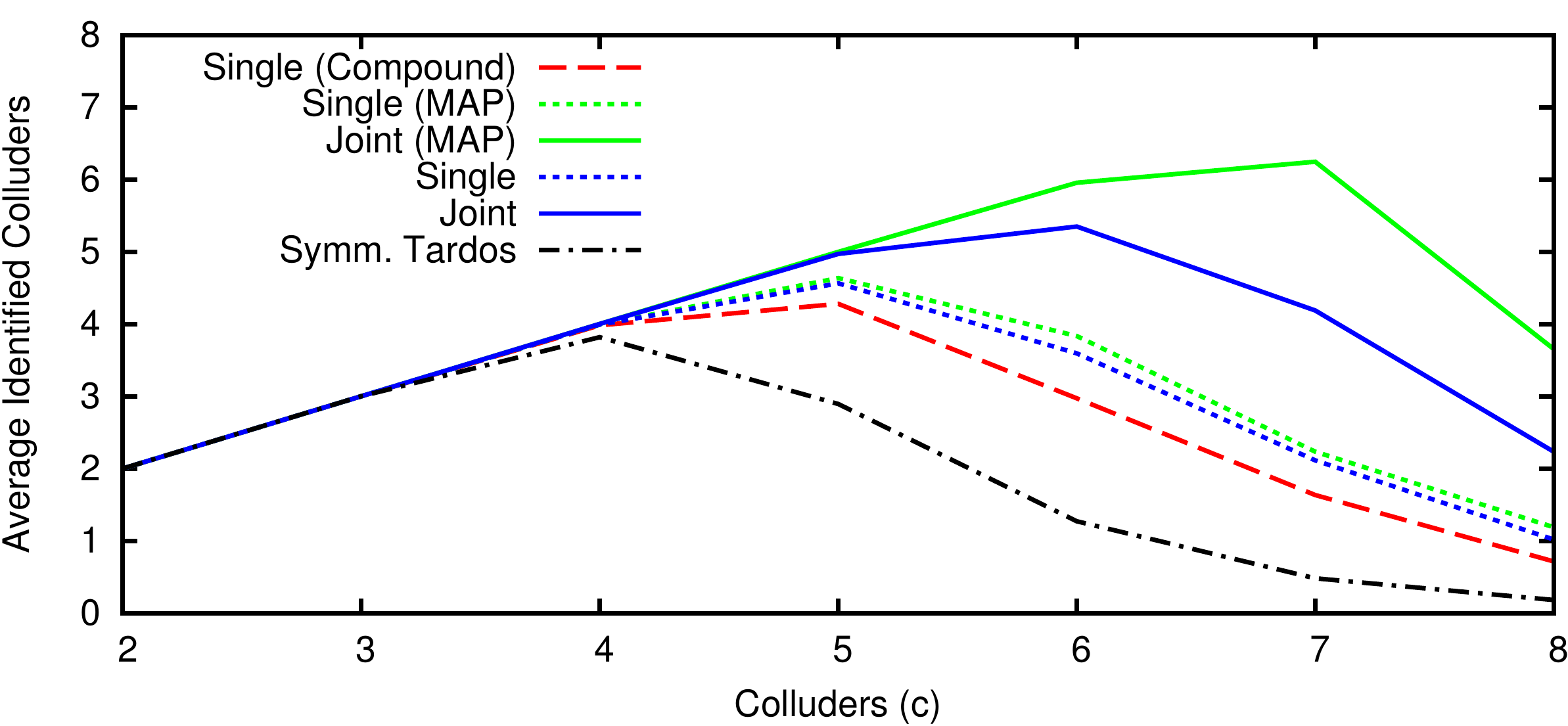}}
\subfigure[\textit{Majority} Attack\label{fig:many-maj}]{\includegraphics[width=0.6\columnwidth]{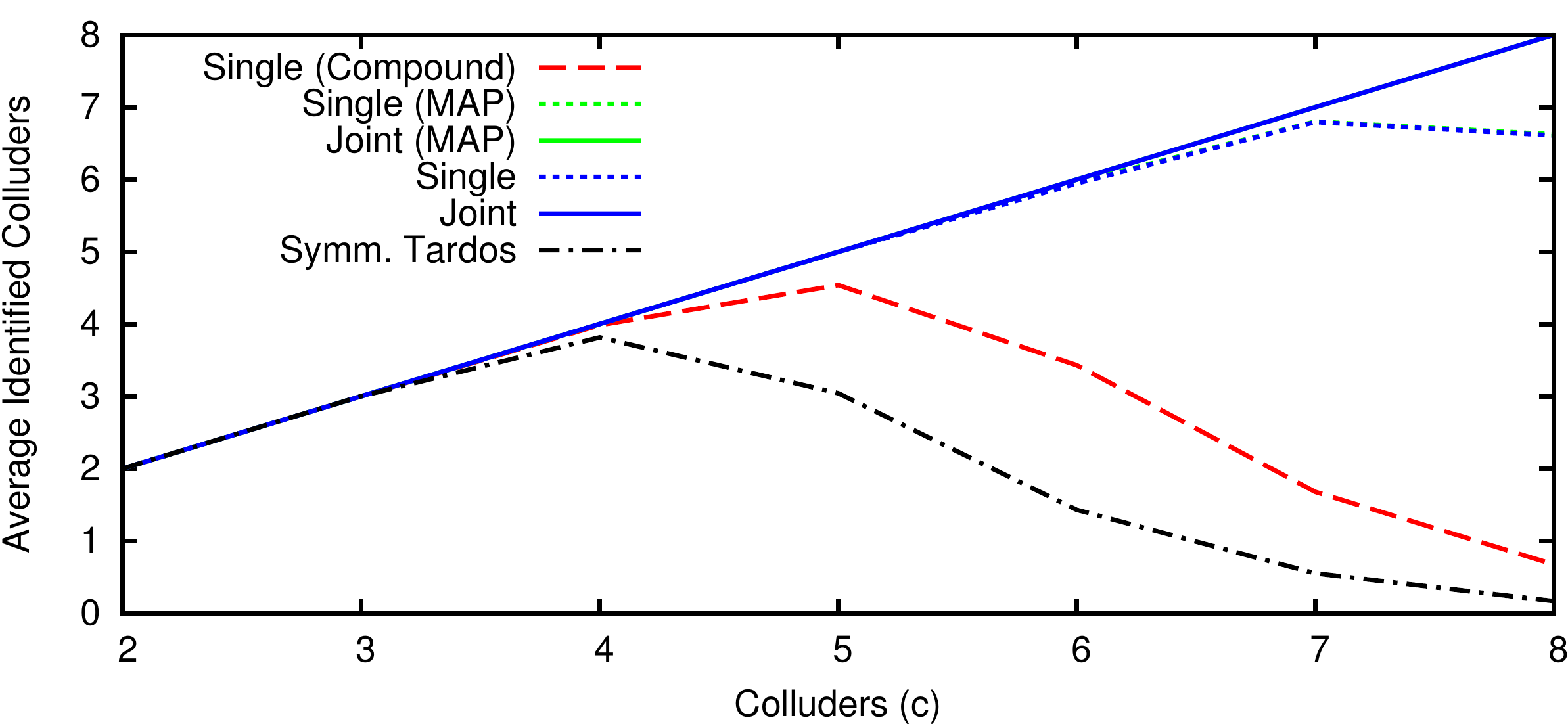}}
\end{center}
\caption{Decoder comparison in the detect-many tracing scenario: $n=10^6$, $m=2048$, $\pfp=10^{-3}$, $\cmax=8$. (Best viewed in color.)\label{fig:catch-many}}
\end{figure}

Figures~\ref{fig:many-wca}--\ref{fig:many-maj} show the average number of identified colluders by different decoding approaches. The experimental setup considers $n=10^{6}$ users, code length $m = 2048$, and several collusion attacks (\textit{worst-case} attacks, i.e. minimizing the achievable rate of a single or joint decoder, \textit{interleaving} and \textit{majority} which is a rather mild attack) carried out by two to eight colluders. The global probability of a false positive error is fixed to $\pfp=10^{-3}$. 

As expected, the MAP single decoder knowing $\thetab_c$ provides the best decoding performance amongst the single decoders, yet is unobtainable in practice.
The symmetric Tardos decoder performs poorly but evenly against all attacks; the single decoder based on the compound channel~\eqref{eq:compound} improves the results only slightly.

The \textit{joint} decoders consistently achieve to identify most colluders -- with a dramatic margin in case the traitors choose the worst-case attack against a single decoder.
This attack bothers the very first step of our decoder, but as soon as some side information is available or a joint decoder is used, this is no longer the worst case attack. Finding the worst case attack against our iterative decoder is indeed difficult. A good guess is the interleaving attack which is asymptotically the worst case against the joint decoder \cite{Huang:2010fk}. The experiments show that it reduces the performance of the \textit{joint} decoders substantially for large $c$.

The decoder based on the inference $\hat{\thetab}_{\cmax}$ and the true MAP are different when $c$ is lower than $\cmax$. However, this is not a big deal in practice for a fixed $m$: for small $c$, the code is long enough to face the collusion even if the score is less discriminative than the ideal MAP; for big $c$ the score of our decoder gets closer to the ideal MAP.  

\subsection{Decoding performance of the soft decoder}

We assess the performance of the soft decision decoders proposed in Sec.~\ref{sec:soft} in two tracing scenarios: (i) Kuribayashi considers in~\cite{Kuribayashi10e} $n=10^{4}$ users and code length $m=10^{4}$, (ii) a large-scale setup with $33\,554\,432$ users and $m=7\,440$ where Jourdas and Moulin \cite{Jourdas09a} provide results for their high-rate random-like fingerprinting code under averaging and interleaving attack. 

\begin{figure}[t]
\begin{center}
\subfigure[\textit{Worst-Case} Attack against Single Decoder]{\includegraphics[width=0.58\columnwidth]{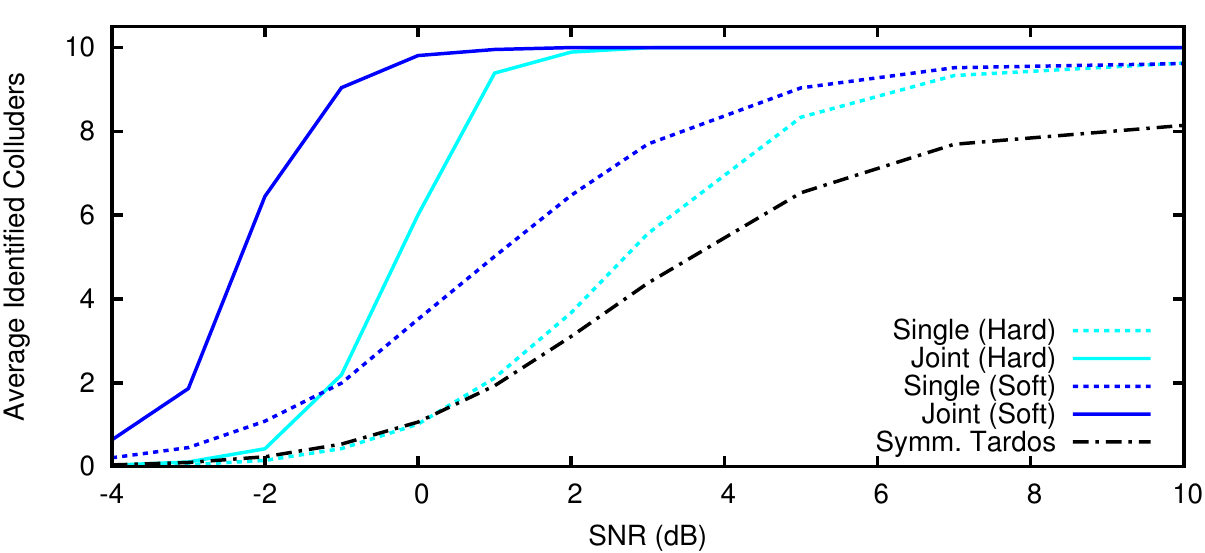}}
\subfigure[\textit{Interleaving} Attack]{\includegraphics[width=0.58\columnwidth]{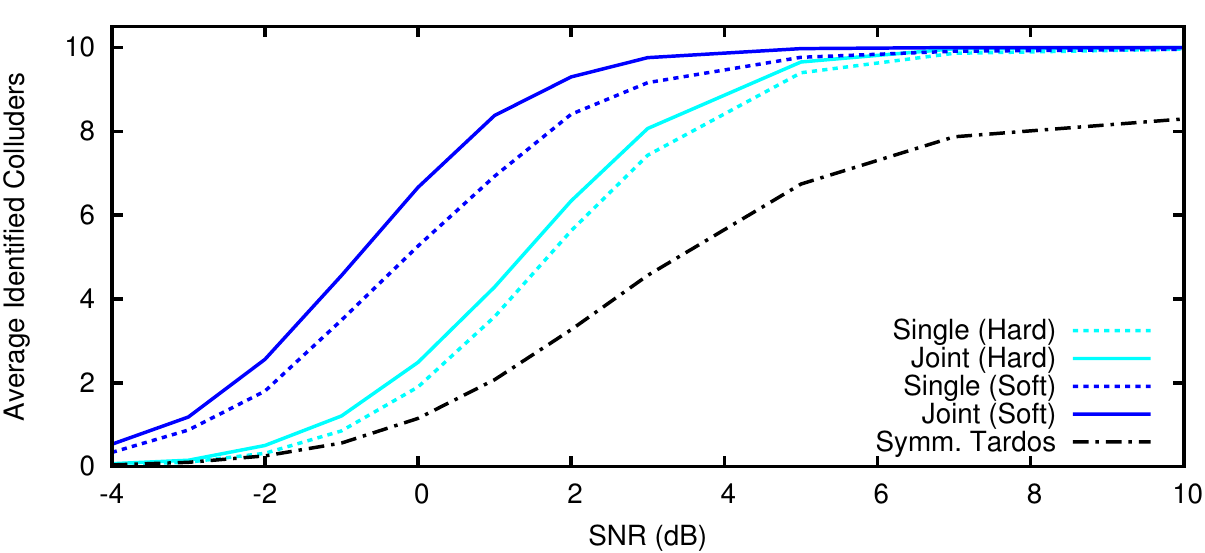}}
\subfigure[\textit{Majority} Attack]{\includegraphics[width=0.58\columnwidth]{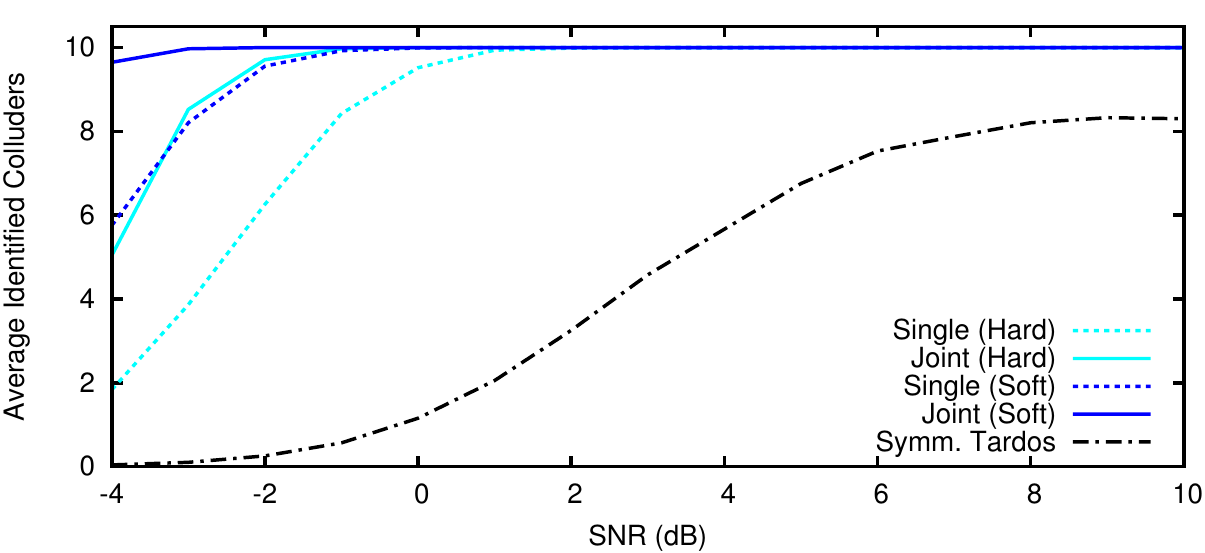}}
\subfigure[\textit{Averaging} Attack]{\includegraphics[width=0.58\columnwidth]{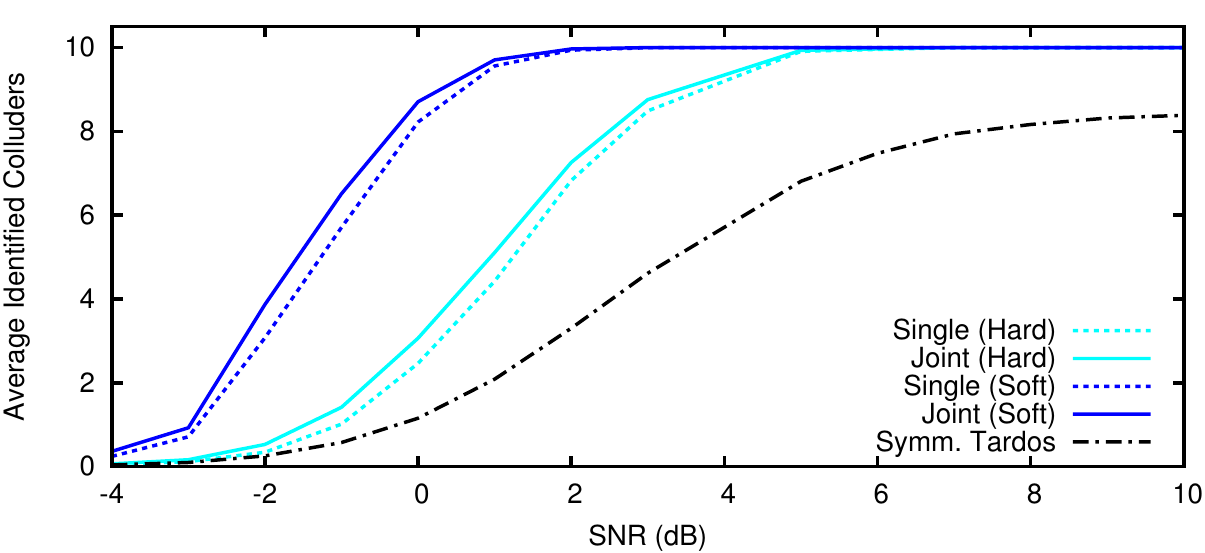}}
\end{center}
\caption{Kuribayashi setup: $n=10^4$, $m=10^4$, $\pfp=10^{-4}$, $c=10$, $\cmax=20$; \textit{worst-case}, \textit{interleaving}, \textit{majority} and \textit{averaging} attack followed by AWGN ($-4,\dots,10$ dB SNR).\label{fig:kuri}}
\end{figure}

In Fig.~\ref{fig:kuri}, we compare the average number of identified colluders for the \textit{single} and \textit{joint} decoder using different estimates of the collusion process: \textit{hard} relates to decoders using hard thresholding and $\hat{\thetab}_{\cmax}$ while \textit{soft} identifies the noise-aware decoders relying on $\hat{\thetab}^{(I)}_{\cmax}$ or $\hat{\thetab}^{(II)}_{\cmax}$ chosen adaptively based on the likelihood of the two models. All plots also show the results for the (hard-thresholding) symmetric Tardos decoder.
The false-positive rate is set to $10^{-4}$. Extensive experiments ($3\cdot10^{4}$ test runs) have been carried out to validate the accusation threshold obtained by rare-event simulation. As expected, soft decoding offers substantial gains in decoding performance. The margin between the \textit{single} and \textit{joint} decoders depends on the collusion strategy. Dramatic improvements can be seen when the collusion chooses the \textit{worst-case} attack against a single decoder, cf. Fig.~\ref{fig:kuri}(a). On the other hand, the gain is negligible when averaging is performed.

Note that the attacks in (a)--(c) pertain to the pick-and paste attacks while Fig.~\ref{fig:kuri}(d) shows the linear \textit{averaging} attack.

Comparison with the results provided in~\cite{Kuribayashi10e} for the \textit{majority} attack is difficult: (i) they were obtained for Nuida's discrete code construction~\cite{Nuida09a} tuned on $c=7$ colluders, and (ii) the false-positive rate of~\cite{Kuribayashi10e} does not seem to be under control for the symmetric Tardos code. We suggest to use the \textit{hard} symmetric Tardos decoder~\cite{Skoric2008:Symmetric} as a baseline for performance comparison. By replacing the accusation thresholds proposed in~\cite{Kuribayashi10e} with a rare-event simulation, we are able to fix the false-alarm rate in case of the symmetric Tardos code. Furthermore, the decoding results given in \cite{Kuribayashi10e} for the discrete variant of the fingerprinting code ({\it i.e.} Nuida's construction) could be significantly improved by rare-event simulation based thresholding. Contrary to the claim of~\cite{Kuribayashi10e}, \textit{soft} decision decoding always provides a performance benefit over the \textit{hard} decoders.

\begin{figure}[t]
\begin{center}
\subfigure[\textit{Averaging} Attack\label{fig:mouli-avg}]{\includegraphics[width=0.6\columnwidth]{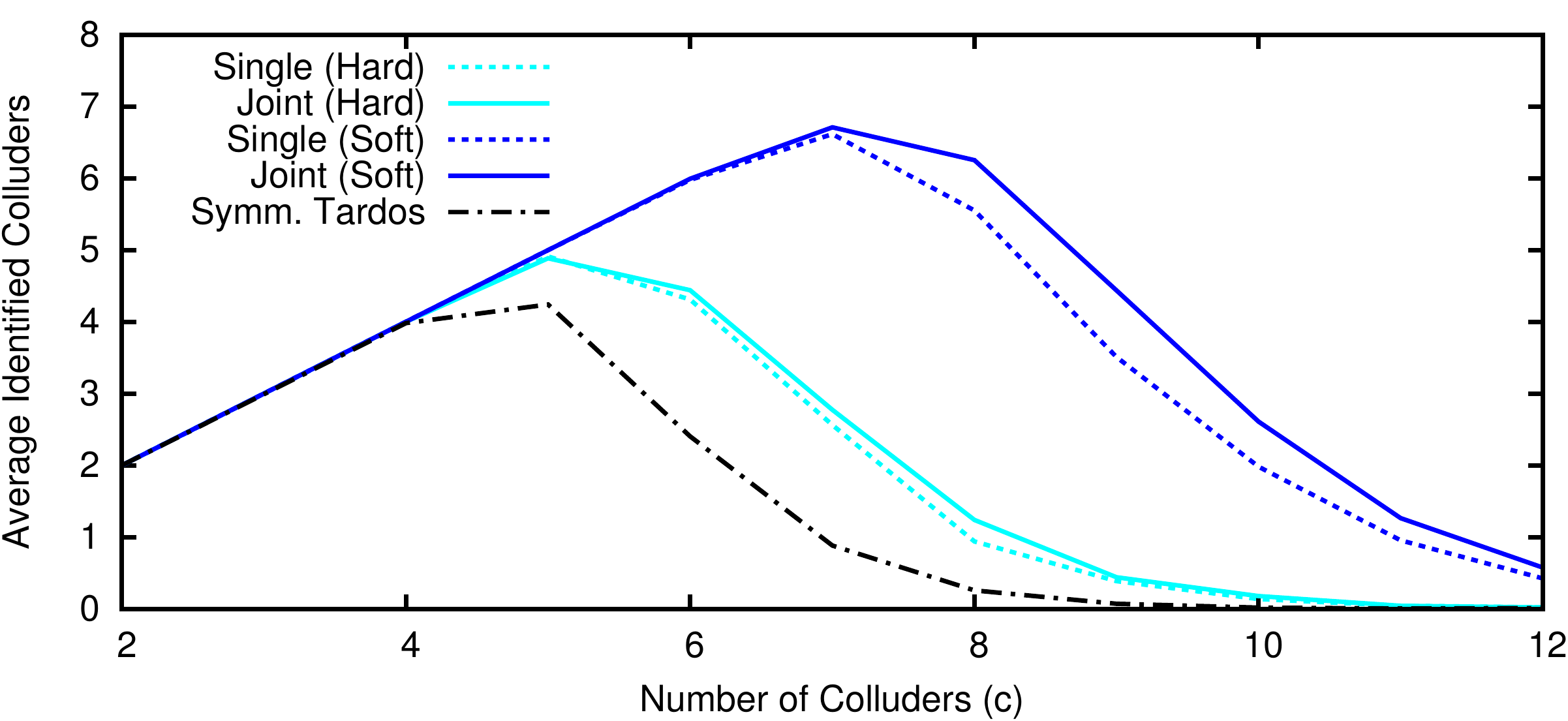}}
\subfigure[\textit{Interleaving} Attack\label{fig:mouli-int}]{\includegraphics[width=0.6\columnwidth]{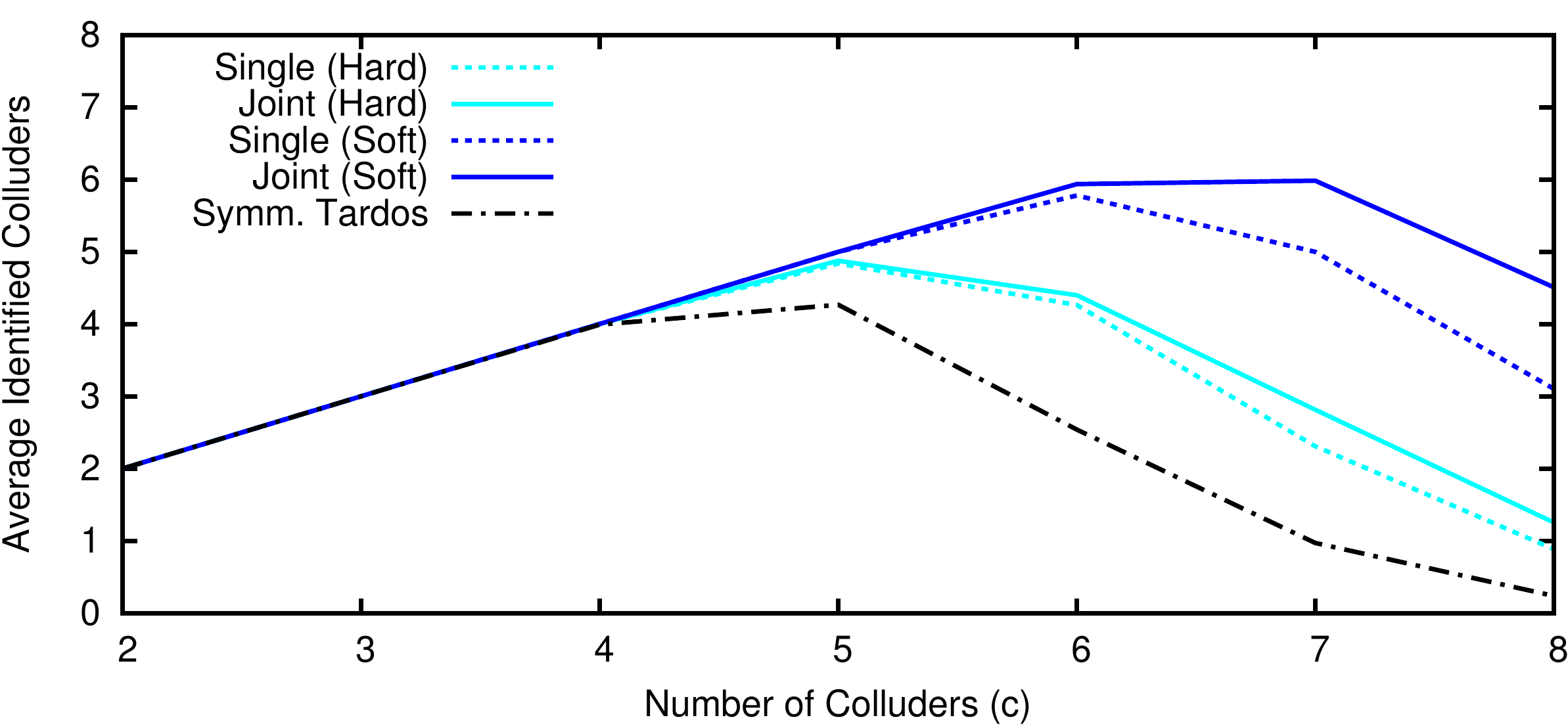}}
\end{center}
\caption{Jourdas \& Moulin setup: $n=33\,554\,432$, $m=7\,440$, $\pfp=10^{-3}$, \textit{averaging} and \textit{interleaving} attack followed by AWGN ($0$ dB SNR).\label{fig:mouli}}
\end{figure}

In Fig.~\ref{fig:mouli} we illustrate the decoding performance when dealing with a large user base. We consider \textit{averaging} and \textit{interleaving} attacks by $c=2,\dots,12$ and $c=2,\dots,8$ colluders ($\cmax=12$ and $\cmax=8$, respectively) followed by AWGN with variance $\sn=1$. The global false-positive rate is set up to $10^{-3}$. The benefit of the \textit{soft} decoding approach in clearly evident. Joint decoding provides only a very limited increase in the number of identified colluders. For comparison, Jourdas \& Moulin indicate an error rate of $\pe=0.0074$ for $c=10$ colluders in the first, and $\pe=0.004$ for $c=5$ colluders in the second setting for a detect-one scenario~\cite{Jourdas09a}. 

In \cite{Jourdas08a}, $\pfp=0.0016$ and $\pfn=0.044$ are given for the first experiment (Fig.~\ref{fig:mouli-avg}) by introducing a threshold to control the false-positive rate. Our \textit{soft} \textit{joint} decoder achieves a $\pfn=0.046$ for $\pfp=10^{-3}$ (for $c=10$ colluders), catching $2.6$ traitors on average.

In the second experiment (see Fig.~\ref{fig:mouli-int}), our \textit{joint} decoder compares more favorably: with the given code length, all $c=5$ colluders can be identified and for a collusion size $c=8$, $4.5$ traitors are accused without observing any decoding failure in $3\cdot10^{3}$ tests.

\subsection{Runtime Analysis}

Single decoding can be efficiently implemented to compute more than one million scores for a code of length $m=1024$ per second. Its complexity is in $O(n \cdot m)$.
Selecting the $\pt$ most likely guilty users can be efficiently done with the max-heap algorithm. Yet, it consumes a substantial parts of the runtime for small $m$. The runtime contribution of the joint decoding stage clearly depends on the size of pruned list of suspects, $O(m \cdot \pt)$ and is independent of the subset size $t$ thanks to the \textit{revolving door} enumeration method of the subsets\footnote{In each step $\varphib$ is updated by replacing one user's codeword. See \cite{Meerwald11c} for details.}. Restricting $\pt$ and $\tmax$ keeps the joint decoding approach computationally tractable. Better decoding performance can be obtained using higher values at the cost of a substantial increase in runtime. Experiments have shown that even the moderate settings ($\pt\approx 4.5\cdot10^{6}$ and $\tmax=5$) achieve a considerable gain of the joint  over the single decoder for several collusion channels.

Thresholding accounts for more than half of the runtime in the experimental setups investigated in this work. However, this is not a serious issue for applications with a large user base or when $\pt$ becomes large. Thresholding depends on the subset size $t$ because a large number of random codeword combinations must be generated and because we seek lower probability level in $O(\pfp/n^{t})$. Therefore, the complexity is in $O(m \cdot t^{2} \cdot \log(n))$ according to~\cite{Guyader:2011kx}. There are no more than $\cmax$ such iteration with $t\leq\cmax$, so that the global complexity of our decoder stays in $O(m\log(n))$.

More details about the runtime are given in~\cite{Meerwald11c}. Note that results have been obtained with a single CPU core although a parallel implementation can be easily achieved. 


\section{Conclusion}
\label{sec:conclusion}
Decoding probabilistic fingerprinting codes in practice means to trace guilty persons over a large set of users while having no information about the size nor the strategy of the collusion. This must be done reliably by guaranteeing a controlled probability of false alarm. 

Our decoder implements provably good concepts of information theory (joint decoding, side information, linear decoder for compound channels) and statistics (estimation of extreme quantile of a rare event). Its extension to soft output decoding is straightforward as its does not change its architecture.

Since the proposed iterative method is neither just a single decoder nor completely a joint decoder (it only considers subsets over a short list of suspects), it is rather difficult to find the best distribution for code construction and its worst case attack.  Experiments show that the interleaving attack is indeed more dangerous than the worst-case attack against a single decoder.

\appendix
\label{appendix}
We prove that $\mathcal{E}_{\cmax}(\thetab_{c})=\{\tilde{\thetab}_{k}| k\leq\cmax, \Prob{y|p,\tilde{\thetab}_{k}}=\Prob{y|p,\thetab_{c}},\,\forall (y,p)\in\{0,1\}\times[0,1]\}$ is one sided. The collusion channels of this set share the property that $\Prob{Y=1|p,\tilde{\thetab}_{k}}=q(p)\geq 0, \forall p\in[0,1]$.
From~\cite[Eq.~(20)]{Furon2009:Worst-WIFS}:
\begin{eqnarray}
\Prob{Y=1|X=1,p,\tilde{\thetab}_{k}}&=&q(p)+k^{-1}(1-p)q'(p)\\
\Prob{Y=1|X=0,p,\tilde{\thetab}_{k}}&=&q(p)-k^{-1}pq'(p)
\end{eqnarray}

Take $(\tilde{\thetab}_{k_{A}},\tilde{\thetab}_{k_{B}})\in\mathcal{E}_{\cmax}(\thetab_{c})^{2}$  s.t. $k_{A}<k_{B}$. We first show that $R(f_{T},\tilde{\thetab}_{k_{A}})>R(f_{T},\tilde{\thetab}_{k_{B}})$ so that the minimizer of $R(f_{T},\thetab)$ over $\mathcal{E}_{\cmax}(\thetab_{c})$ is indeed $\tilde{\thetab}_{\cmax}$. Denote by $(\mu_{1},\mu_{2})$ the following conditional probability distributions:
\begin{eqnarray}
\mu_{1}(y,x|p)&=&\Prob{Y=y|p}=q(p)^{y}(1-q(p))^{(1-y)}\\
\mu_{2}(y,x|p)&=&\Prob{Y=y|X=x,p,\tilde{\thetab}_{k_{A}}}. 
\end{eqnarray}
Then, $\Prob{Y|X,p,\tilde{\thetab}_{k_{B}}} = (1-\lambda) \mu_{1}(Y,X|p) + \lambda\mu_{2}(Y,X|p)$, $\forall p\in[0,1]$, with $\lambda = k_{A}/k_{B}<1$. The mutual information is a convex function of $\Prob{Y|X,p}$ for fixed $\Prob{X|p}$ so that, once integrated over $f_{T}(p)$, we have
\begin{equation}
R(f_{T},\tilde{\thetab}_{k_{B}}) \leq (1-\lambda) \cdot 0 + \lambda \cdot R(f_{T},\tilde{\thetab}_{k_{A}})<R(f_{T},\tilde{\thetab}_{k_{A}}).
\end{equation}

We now prove that~\eqref{eq:DefOneSided} holds $\forall \thetab\in\mathcal{E}_{\cmax}(\thetab_{c})$. This is equivalent to
\begin{equation}
R(f_{T},\tilde{\thetab}_{k})-D(\Prob{Y,X|\tilde{\thetab}_{k}}||\Prob{Y,X|\tilde{\thetab}_{\cmax}})-R(f_{T},\tilde{\thetab}_{\cmax})\geq 0,
\end{equation}
where the LHS is of the form $\Exp{P\sim f_{T}}{g(P)}$. After developing the expressions, we find that:
\begin{eqnarray}
g(p)&=&(k^{-1}-\cmax^{-1})p(1-p) \cdot \nonumber \\
 & & \left(q'(p)\log\left(1+\frac{1-p}{\cmax}\frac{q'(p)}{q(p)}\right)\right. + \nonumber \\
 & & q'(p)\log\left(1+\frac{p}{\cmax}\frac{q'(p)}{1-q(p)}\right) - \nonumber \\
 & & q'(p)\log\left(1-\frac{1-p}{\cmax}\frac{q'(p)}{1-q(p)}\right) - \nonumber \\
 & & \left.q'(p)\log\left(1-\frac{p}{\cmax}\frac{q'(p)}{q(p)}\right)\right)
\end{eqnarray}
The four terms inside parenthesis are not negative because, with $\gamma>0$, $x\log(1+\gamma x)\geq 0$  for $x>-\gamma^{-1}$.  Since $k\leq\cmax$, we obtain $g(p)\geq0$, whence~\eqref{eq:DefOneSided}.

\bibliographystyle{IEEEtran}
\bibliography{fp,impl,group_testing}

%

\end{document}